\documentclass{jfm}

\usepackage{graphicx}
\usepackage{epstopdf,epsfig}

\usepackage[czech,english]{babel}

\usepackage{mathtools}
\usepackage{natbib} 

\newcommand{\dd}{\,\mathrm{d}}
\newcommand{\OBig}{\mathcal{O}}

\newcommand{\rr}{\boldsymbol{r}}
\newcommand{\uu}{\boldsymbol{u}}
\newcommand{\ee}{\boldsymbol{e}}

\newcommand*{\HeFour}{$^4$He}

\newcommand*{\OmegaMF}{\Omega_{0}}  
\newcommand*{\vb}[1]{\boldsymbol{#1}}  
\newcommand*{\vdot}{\vb{\cdot}}     
\newcommand*{\vvec}{\vb{v}}         

\newcommand*{\xp}{\vb{x}_\text{p}}  
\newcommand*{\vp}{\vb{v}_\text{p}}  
\newcommand*{\StokesTime}{\tau_\text{p}}
\newcommand*{\rhop}{\rho_\text{p}}
\newcommand*{\ap}{a_\text{p}}    

\newcommand*{\vn}{\vvec_{\text{n}}}
\newcommand*{\vs}{\vvec_{\text{s}}}

\newcommand*{\pn}{p_{\text{n}}}
\newcommand*{\ps}{p_{\text{s}}}
\newcommand*{\rhon}{\rho_{\text{n}}}
\newcommand*{\rhos}{\rho_{\text{s}}}
\newcommand*{\nun}{\nu_{\text{n}}}
\newcommand*{\nus}{\nu_{\text{s}}}

\newcommand*{\Fns}{\vb{f}_{\text{ns}}}

\newcommand*{\gradient}{\vb{\nabla}}
\newcommand*{\laplacian}{\nabla^2}
\newcommand*{\diffp}[2]{\frac{\partial #1}{\partial #2}}
\newcommand*{\diff}[2]{\frac{\mathrm{d} #1}{\mathrm{d} #2}}
\newcommand*{\Diff}[2]{\frac{\mathrm{D} #1}{\mathrm{D} #2}}

\shorttitle{On the determination of vorticity using Lagrangian particles}
\shortauthor{O.~Outrata et al.}

\title{On the determination of vortex ring vorticity using Lagrangian particles}

\author{O.~Outrata\aff{1},
  M.~Pavelka\aff{1},
  J.~Hron\aff{1},
  M.~La~Mantia\aff{1}
  \corresp{\email{lamantia@mbox.troja.mff.cuni.cz}},
  J.~I.~Polanco\aff{2}
  and G.~Krstulovic\aff{2}}

\affiliation{\aff{1}Faculty of Mathematics and Physics, Charles University, Ke Karlovu 3, 121\,16 Prague, Czech Republic
\aff{2}Universit\'{e} C\^{o}te d'Azur, Observatoire de la C\^{o}te d'Azur, CNRS, Laboratoire Lagrange, Boulevard de l'Observatoire CS 34229 -- F 06304 Nice Cedex 4, France}

\begin{document}

\maketitle

\begin{abstract}
Particles are a widespread tool for obtaining information from fluid flows.
When Eulerian data are unavailable, they may be employed to estimate flow fields or to identify coherent flow structures.
Here we numerically examine the possibility of using particles to capture the dynamics of isolated vortex rings propagating in a quiescent fluid.
The analysis is performed starting from numerical simulations of the Navier--Stokes and the Hall--Vinen--Bekarevich--Khalatnikov equations, respectively describing the dynamics of a Newtonian fluid and a finite-temperature superfluid.
The flow-induced positions and velocities of particles suspended in the fluid are specifically used to compute the Lagrangian pseudovorticity field, a proxy for the Eulerian vorticity field recently employed in the context of superfluid $^4$He experiments.
We show that, when calculated from ideal Lagrangian tracers or from particles with low inertia, the pseudovorticity field can be accurately used to estimate the propagation velocity and the growth of isolated vortex rings, although the quantitative reconstruction of the corresponding vorticity fields remains challenging.
On the other hand, particles with high inertia tend to preferentially sample specific flow regions, resulting in biased pseudovorticity fields that pollute the estimation of the vortex ring properties.
Overall, this work neatly demonstrates that the Lagrangian pseudovorticity is a valuable tool for estimating the strength of macroscopic vortical structures in the absence of Eulerian data, which is, for example, the case of superfluid $^4$He experiments.
\end{abstract}



\section{Introduction}
\label{sec:intro}

The accurate and well-resolved measurement in space and time of fluid flows represents one of the main ongoing challenges in experimental fluid dynamics.
A number of methods have been developed to capture the evolution of velocity fields and other relevant quantities in laboratory flows.
A common approach is to visualize the motion of relatively small particles advected by the flow of interest, from which properties of the latter can be inferred \citep[see, for example,][]{piv}.
Various techniques can then be employed to quantitatively process the obtained images and the choice mainly depends on the features to be investigated.

The most popular tools are particle image velocimetry (PIV), which allows to access instantaneous flows fields, i.e. Eulerian data, and particle tracking velocimetry (PTV), which is a Lagrangian
technique, focusing on the temporal dynamics of the visualized particles.
In principle, both methods can be used to gather complementary information on the same flow but, in practice, this is not always possible.
Leaving aside for a moment the important roles played by particle size and inertia, addressed below, the main limiting factor is the particle number density, discussed, for example, by \citet{holger}.
Specifically, if the number of particles per unit area is smaller than about $0.01$~pixel$^{-2}$, significant errors occur in PIV measurements.
It then follows that, if one cannot seed the flow of interest with enough particles, PTV is the only choice to get quantitative information from the collected images.

An exemplary case is represented by superfluid $^4$He, which is also known as He~II \citep[see, for example,][]{bss,mjs}.
The cryogenic flows of this unique liquid are routinely visualized and several PTV studies have been reported to date -- see \citet{pato21} for a recent example.
However, at a given time, one can usually identify the positions of about 100 particles within a 1~Mpixel image, each particle having an apparent size of a few pixels.
Consequently, the visualization data presently collected in He~II do not match the quality of the PIV data obtained in water or air, i.e. the former data are more suitable for PTV than for PIV, at least if one has in mind quantitative studies.

The observed behaviour is related to the available experimental techniques, which do not allow to seed the flow of interest with a relatively large number of particles, at a given time.
The density of the liquid is equal to about 145~kg~m$^{-3}$ \citep{donnelly98} and it is consequently challenging to employ neutrally buoyant particles, as discussed, for example, by \citet{prf}.
Indeed, the mass density of the flow-probing particles should match that of the liquid but this is yet to be achieved for visualization experiments in He~II.
It follows that the particles routinely used -- made of solid hydrogen or deuterium -- disappear from the field of view after some time, of the order of minutes.
Additionally, particles are usually injected into the liquid and, before taking measurements, one must wait for the injection flow to disappear.
Then, as already said, one is left with about 100 particles that can be tracked for some time, of the order of seconds (the particle size is usually of the order of micrometers).

\begin{figure}
  \centerline{\includegraphics[width = 0.7\linewidth]{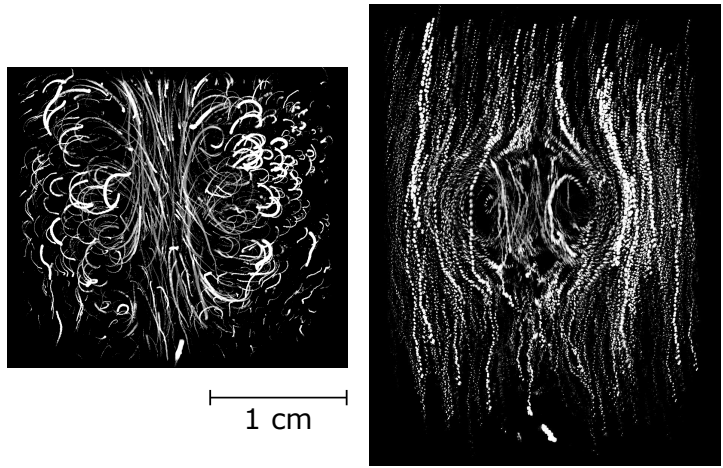}}
  \caption{Visualisation of particle trajectories induced by a vortex ring in superfluid $^4$He at $1.50$~K, with Reynolds number $\Rey \approx 10^5$; see \citet{pato20} for experimental details.
  The white regions correspond to particles reflecting incoming light and the images were obtained from averaging the light intensity reflected by the particles over hundreds of movie frames.
  In the left panel -- $25$~mm wide and $22$~mm high -- the nozzle ejecting the fluid parcel is located at the bottom centre of the image and the vortex ring moves upward.
  The right panel displays (to scale) the same light intensity values of the left one in the coordinate system moving with the vortex ring, where the latter time-dependent position and velocity were computed as discussed by \citet{pato20}.}
 \label{holes}
\end{figure}

To illustrate these facts, we display in figure~\ref{holes} two images obtained from experimental data discussed by \citet{pato20}.
Macroscopic vortex rings propagating in superfluid $^4$He were studied by using the PTV technique, i.e. solid deuterium particles were dispersed in the liquid and illuminated by a planar laser sheet.
The ring-induced motions of the particles were collected by a digital camera, in a plane approximately parallel to the ring propagation direction, that is, in a plane approximately corresponding to one of the ring symmetry planes.
The white regions correspond to particles reflecting incoming light and the images were obtained by superimposing hundreds of frames, i.e. the white regions can be viewed as preliminary indications of the tracks followed by the particles (heavier than the fluid in this case).
The left panel corresponds to the superposition of movie frames in the laboratory coordinate system, whereas, in the right one, the white regions have been centred in the reference system moving with the vortex ring.

We clearly appreciate the emergence of holes, which would not be observed for perfect tracers distributed uniformly through the experimental domain.
Indeed, even for relatively small particles, when their action on the flow can be neglected, particle inertia might lead to the decoupling of the particle trajectories from the flow streamlines and, in particular, to preferential concentration \citep[see, for example,][]{Maxey1987,Balachandar2010}.

Additionally, in comparison to typical PIV data, very few particles are seen in figure~\ref{holes}, that is, these data cannot be employed for a robust estimate of Eulerian quantities, such as velocity and vorticity fields.
Indeed, in PTV experiments, one does not usually have access to the vorticity field because the latter requires a very large number of particles, vorticity being a small-scale quantity that can present sharp variations in time and space.
Nevertheless, having access to the vorticity field -- or a proxy thereof -- can be of great interest for the description of laboratory flows, e.g. for quantitative comparisons between particle trajectories obtained in different experimental conditions.

This research route was specifically followed by \citet{pato20} in their PTV study on the propagation of macroscopic vortex rings in He~II.
The ring-induced motions of the particles were employed to calculate the Lagrangian pseudovorticity field -- a proxy for the Eulerian vorticity field, which can be used to quantitatively measure the strength of the generated vortex rings and which can also provide information on the size, position and velocity of these objects.

Additionally, \citet{pato20} have shown that, in conditions yet to be met in experiments, the Lagrangian pseudovorticity $\theta$ is equal to half of the the Eulerian vorticity $\omega$.
The present work further clarifies the relation between $\theta$ and $\omega$, especially in conditions closer to those encountered in typical experiments.
We not only consider in the following how particle concentration affects this relation but also take into account the role of particle inertia.

To this end, we numerically study the motion of particles induced by vortex rings in two different settings.
In the first part of the work, we investigate the behaviour of an isolated vortex ring propagating in a Newtonian fluid by using the Finite Element Method (FEM) \citep{fem} with the Incremental Pressure Correction Scheme \citep{ipcs}.
We then compute the Lagrangian pseudovorticity field from the positions and velocities of fluid particles (tracers) advected by the obtained Eulerian flow fields, as a function of experimentally relevant parameters, such as the number of flow-probing particles and their spatial distribution.

Secondly, we investigate the effect of particle inertia on the relation between pseudovorticity and vorticity, in the specific case of a vortex ring propagating in a superfluid.
For this, we perform direct numerical simulations of the Hall--Vinen--Bekarevich--Khalatnikov (HVBK) model, which is often employed to describe the large-scale hydrodynamics of He~II.
Indeed, the dynamics of relatively small spherical particles in flows of superfluid $^4$He has been recently investigated within this theoretical framework \citep{nacho20}.

In summary, the results presented below reinforce the view that the Lagrangian pseudovorticity can be used to quantitatively estimate the strength of relatively large vortical structures in the absence of Eulerian data, which is specifically the case of experiments in He~II.
We confirm that the magnitude of $\theta$ can be appreciably smaller than that of $\omega$, consistently with previous analytical and experimental estimates \citep{pato20}.
We also show that, in order to get a closer quantitative agreement between the vorticity and pseudovorticity trends, larger particle concentrations should be employed compared to those commonly used in experiments.
Furthermore, our results demonstrate that the Lagrangian pseudovorticity can be used to resolve accurately general features of the observed ring dynamics.
Such features include the ring time-dependent position and propagation velocity, especially in the case of tracers and particles with sufficiently low inertia.
However, we also report that, at present, $\theta$ does not seem suitable for capturing the fine details of the vortex ring structure, that is, for reconstructing the corresponding Eulerian vorticity fields.


\section{Lagrangian pseudovorticity}
\label{sec:pseudo}

The vortices shed at the edges of relatively large objects oscillating in superfluid $^4$He have been recently visualised in planes approximately parallel to the direction of oscillation, by following the flow-induced motions of small solid particles suspended in the liquid \citep{duda15,duda17}.
The strength of these vortices was estimated from the two-dimensional particle positions and velocities by using a purpose-made scalar measure, introduced to quantitatively compare vortices obtained in different experimental conditions because, as already noted, it is in general not possible to obtain flow vorticity fields from sparse Lagrangian data.

This scalar quantity was named Lagrangian pseudovorticity $\theta$ and, as mentioned above, it was also used by \citet{pato20} in their experimental study on macroscopic vortex rings propagating in superfluid $^4$He.
It is here defined as
\begin{equation}
  \theta(\rr,t) = \left\langle \frac{\left[\left(\rr_i - \rr\right) \times \uu_i\right]_z}{|\rr_i - \rr|^2} \right\rangle_\mathcal{M},
  \label{ET}
\end{equation}
where $\rr_i$ and $\uu_i$ denote the particle positions and velocities in the considered plane, respectively.
The angle brackets indicate the ensemble average within the set $\mathcal{M}$ of Lagrangian particles, which are captured within a time window centred at time $t$ and are found within an annular region centered, on a chosen grid, at the inspection point $\rr$.
The subscript $z$ denotes the axis perpendicular to the observation plane, i.e. we consider here the only non-zero component of the vector product (the observed particle tracks occur in a plane).
The size of the annular region was chosen in order to have enough particles for the calculation of $\theta$ and, at the same time, to exclude diverging contributions from particles too close to the inspection point $\rr$.

At this point, it is useful to note a few properties of the Lagrangian pseudovorticity $\theta$.
As mentioned above, it can be shown analytically that $\theta$ is closely related to the Eulerian vorticity, in the case of fluid particles that are homogeneously distributed in space \citep{pato20}.
Furthermore, it was reported that, for a Rankine vortex, the magnitude of $\theta$ decreases when the area of the annular region used for its calculation increases, that is, when the outer radius of the region increases, for given grid and number of fluid particles within the annular region \citep{pato20}.
Additionally, this magnitude can be appreciably smaller than that of $\omega$ when the area is significantly larger than the area where the flow vorticity is concentrated.
For example, when the region outer radius is 10 (100) times larger than the Rankine vortex radius, the maximum pseudovorticity magnitude, located at the vortex centre, is more than 5 (50) times smaller than the corresponding vorticity value.
Note in passing that changing the grid or the number of points within the annular region has less significant effect on the calculated pseudovorticity magnitude, in comparison with the just mentioned influence of the annular region size.

\begin{figure}
  \centerline{\includegraphics[width = 0.7\linewidth]{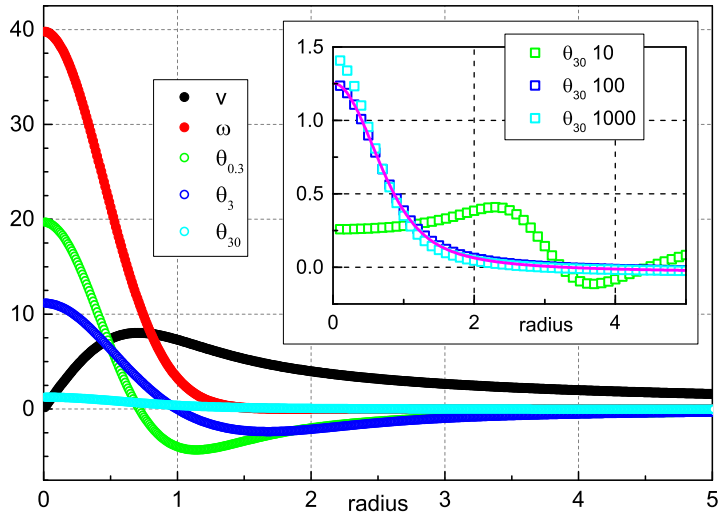}}
  \caption{Pseudovorticity for a Lamb-Oseen vortex as a function of the distance from the vortex axis (radius).
  Vortex parameters: circulation $\Gamma = 50$~mm$^2$~s$^{-1}$, kinematic viscosity $\nu = 0.01$~mm$^2$~s$^{-1}$ and time $t = 10$~s.
  Black circles: vortex velocity $v$; red circles: vorticity $\omega$; open circles: pseudovorticity $\theta$, the corresponding subscripts in the legend indicate the radius, in mm, of the area used for the calculation of $\theta$ from (\ref{ET}).
  1000 equally spaced grid points were employed for the estimate shown in the main panel, up to a radius of 10~mm, and 100 fluid particles were distributed in the annular region, equally spaced along the radial direction (the pseudovorticity is not computed in the region centre and is set to zero at the vortex centre).
  Note that the chosen particle distribution is not radially symmetric, that is, the particles are clustered in the vortex centre vicinity.
  The results plotted in the inset were obtained by using 100 equally spaced grid points, up to the radius of 10~mm, and the number of equally spaced fluid particles within the annular region is specified in the legend.
  The region outer radius for the data in the inset is 30~mm.
  The magenta line indicates the pseudovorticity trend obtained by using 1000 equally spaced grid points, corresponding to the cyan open circles in the main panel.}
 \label{oseen}
\end{figure}

A similar outcome -- displayed in figure~\ref{oseen} -- can be obtained for a Lamb-Oseen vortex, which is characterised by a vorticity spatial distribution closer to that observed for vortex rings, see, for example, figure~\ref{omaps} in \S\ref{ssec:fem:pseudo}, where relevant numerical results are plotted.
It is also apparent that the pseudovorticity may change sign in the vortex vicinity, similarly to what is seen in the middle panel of figure~\ref{omaps}.
Additionally, it is evident from the inset of figure~\ref{oseen} that the number of fluid particles distributed in the annular region becomes important solely when it assumes relatively small values.
The same applies to the chosen grid size that affects the computed pseudovorticity values only for numbers of grid points smaller than those considered in the figure.

Before presenting our numerical results, we also note here that in appendix \S\ref{sec:anal}, additional analytical results on the relation between pseudovorticity and vorticity are reported.
Specifically, we show that, for an isolated vortex having a purely azimuthal velocity field, the pseudovorticity magnitude becomes smaller, if the area employed for its estimate increases.
We also suggest an explanation for the sign changes of the pseudovorticity displayed in figures \ref{oseen} and \ref{omaps}, that is, the outcome can be explicitly related to the fact that the corresponding particle distributions are not homogenous nor isotropic.


\section{FEM ring}
\label{sec:fem}

The Finite Element Method (FEM) \citep{fem} was chosen for the space discretisation of the well-known equations governing the isothermal dynamics of an incompressible viscous fluid.
We specifically employed the Taylor-Hood mixed FEM by using the FEniCS library \citep{fenics}.
The discretisation in time was performed by employing the implicit Crank-Nicolson scheme and the Incremental Pressure Correction Scheme \citep{ipcs} was used for finding an efficient approximate solution of the coupled velocity-pressure problem.

\begin{figure}
  \centerline{\includegraphics[width = 0.2\linewidth]{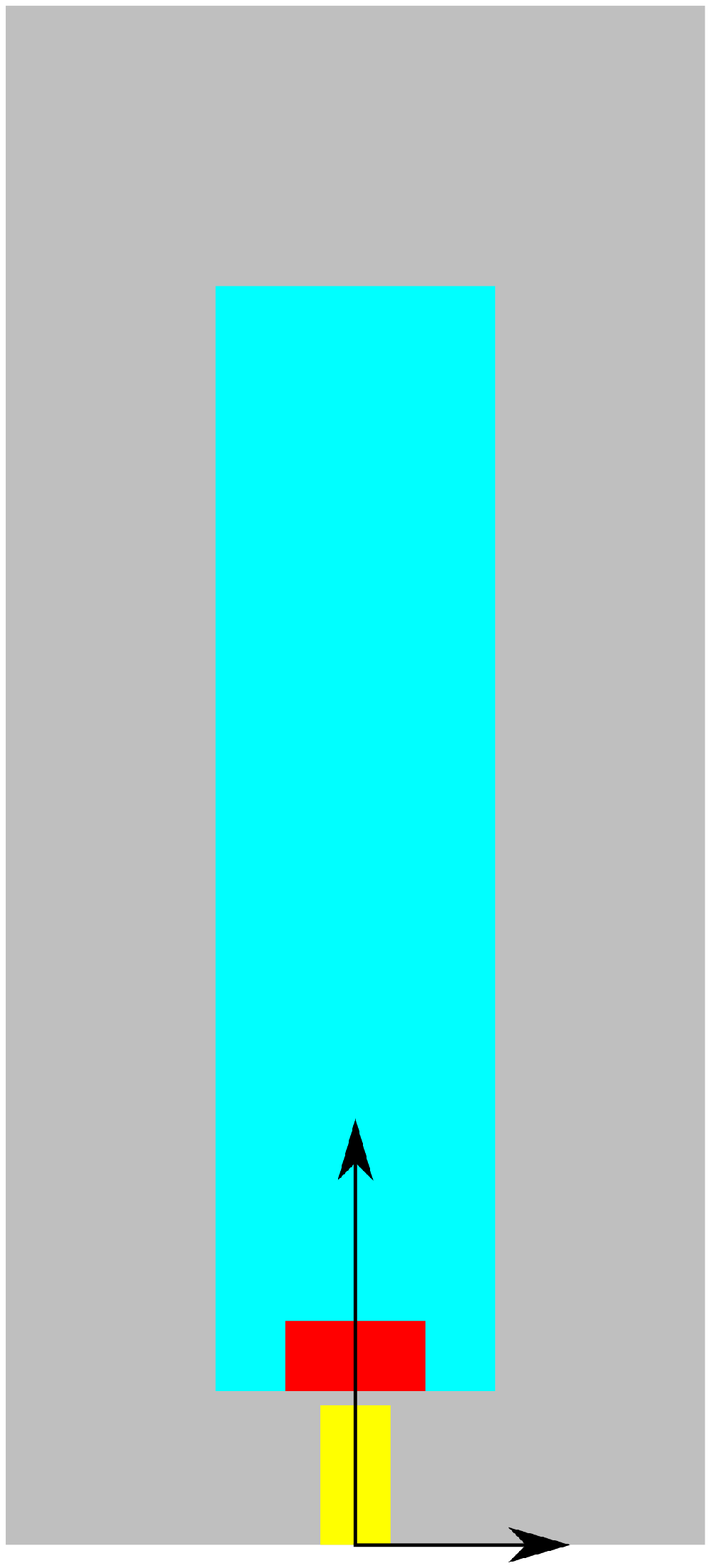}}
  \caption{Two-dimensional domain used for the FEM simulations, composed of the grey, light blue and red areas.
  The yellow area -- 5~mm wide and 10~mm high -- represents the nozzle and is not included in the numerical domain (a parcel of fluid is ejected through the nozzle to generate the vortex ring).
  The grey area is 50~mm wide and 110~mm high.
  In the light blue area -- 20~mm wide and 79~mm high -- 5676 fluid particles are uniformly distributed at the beginning of the simulation (\emph{experiment} particle distribution, labelled E; the area is located 1~mm above the nozzle and includes the red area).
  In the red area -- 10~mm wide and 5~mm high -- 22801 fluid particles are uniformly distributed 1~s after the beginning of the simulation (\emph{delay} particle distribution, labelled D; the area is located 1~mm above the nozzle).
  The origin of the Cartesian reference system (indicated by black arrows) is at the bottom centre of the yellow area.}
  \label{domain}
\end{figure}

A schematic view of the numerical domain employed for the simulations in two dimensions is shown in figure~\ref{domain}.
For these simulations we have specifically chosen boundary conditions close to experimental ones \citep{pato20}.
On all domain boundaries the fluid velocity is set to zero, excluding the top of the domain, where a traction free boundary condition is imposed, and the nozzle top, where the vertical fluid velocity has a spatial parabolic profile during the first second of the simulation (it is zero at later times).
There, the fluid peak velocity linearly increases from 0 to 10~mm~s$^{-1}$ during the first 0.1~s, remains constant for the following 0.8~s and linearly decreases from 10 to 0~mm~s$^{-1}$ during the last 0.1~s of the simulation first second.

The latter boundary condition simulates the vortex ring generation process, which may occur when a parcel of fluid is ejected into a reservoir through a nozzle.
In practice, this is often achieved by using a piston, moving inside the nozzle with a velocity $U_p$, for a stroke length $L$.
An adequate Reynolds number can then be defined as
\begin{equation}
  \Rey = \frac{U_p L}{2 \nu} = \frac{\Gamma_0}{\nu},
  \label{Ere}
\end{equation}
where, for the present simulations, the fluid kinematic viscosity is $\nu = 0.01$~mm$^2$~s$^{-1}$ and $\Gamma_0 = U_p L / 2$ indicates the ring nominal circulation \citep[see, for example,][]{pato20}.

Here, we set $U_p$ equal to the mean value of the imposed fluid velocity during the simulation first second, i.e. $U_p = 6$~mm~s$^{-1}$, and the stroke length $L = U_p t_p = 6$~mm, where $t_p = 1$~s is the imposed duration of the fluid ejection into the reservoir.
It follows that $\Gamma_0 = 18$~mm$^2$~s$^{-1}$ and $\Rey = 1800$, which means that our FEM ring is most likely in the laminar regime \citep{glezer88}.
Additionally, $L/D = 1.2$, where $D = 5$~mm is the nozzle internal diameter, that is, we do not expected a prominent ring wake \citep{gharib98}; note also that the ring nominal impulse, per unit of density, $I_0 = \Gamma_0 \pi (D/2)^2 = 353$~mm$^4$~s$^{-1}$ \citep{sullivan08}.

A mesh independence check was performed on three meshes with different densities for the space discretisation, resulting in systems with sizes ranging from $2 \times 10^5$ to $1.3 \times 10^6$ degrees of freedom.
The size of the time step ranged from 0.01 to 0.001~s.
Full three-dimensional computations were also performed for the sake of comparison and the obtained results were not appreciably different from the two-dimensional ones; see \citet{outrata} for further details.
In the following, we discuss the two-dimensional numerical results obtained by using the most refined mesh, that with approximately $1.3 \times 10^6$ degrees of freedom, and with the smallest time step, equal to 0.001~s (the entire simulation lasts 15~s).

\subsection{Initial fluid particle distributions}
\label{ssec:fem:particles}

We considered fluid particles in the obtained Eulerian flow fields and tracked their motions in order to compute -- from the particle positions and velocities -- the Lagrangian pseudovorticity, (\ref{ET}).
As a first step, we uniformly distributed, at the beginning of the simulation, approximately 6000 particles in a relatively large area of the numerical domain, shown as the light blue region in figure~\ref{domain}, our aim being to replicate actual experimental conditions \citep{pato20}.
We refer to this arrangement as the \emph{experiment} particle distribution and label it with the letter E.
As shown in the following, its results are characterised by the fact that many particles do not move, being away from the vortex ring, similarly to what is observed in experiments.

For our analysis we also employed another particle arrangement and refer to it as the \emph{delay} distribution, because the fluid particles were added to the Eulerian flow fields 1~s after the beginning of the simulation (we label this distribution with the letter D).
Additionally, compared to E, more particles were distributed in a smaller area, shown as the red region in figure~\ref{domain}, our main aim being to increase the number of particles significantly contributing to the pseudovorticity computation, that is, the number of moving particles.

The inner and outer radii of the annular region employed for pseudovorticity estimates, (\ref{ET}), were set to 0.1~mm and 3~mm, respectively, and the corresponding time window is 10~ms wide.
For both distributions, the chosen grid is a rectangular mesh of $53 \times 126$ inspection points, covering a 20.8~mm wide and 60.0~mm high area of the computational domain, centred just above the nozzle, see figure~\ref{domain}.
Note that on the same grid the obtained flow vorticity was interpolated and that the fluid particle velocities are known from the numerical simulation, that is, they are not computed from the particle positions as in the experiments \citep{pato20}.

\subsection{Pseudovorticity maps}
\label{ssec:fem:pseudo}

\begin{figure}
  \centerline{\includegraphics[width = 1\linewidth]{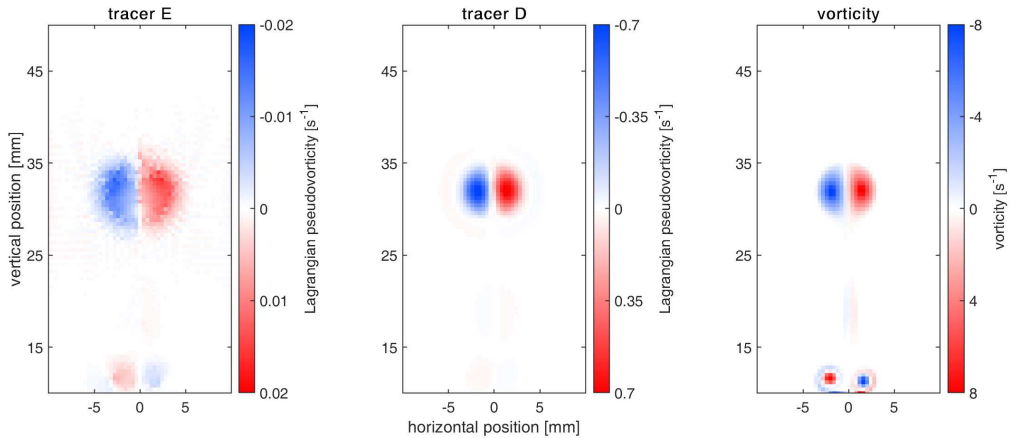}}
  \caption{Pseudovorticity and vorticity maps obtained 10~s after the beginning of the FEM simulation.
  Left and middle panels: pseudovorticity computed from the \emph{experiment} (E) and \emph{delay} (D) particle distributions, respectively.
  Right panel: vorticity field.
  The reference system is that defined in figure~\ref{domain}.
  Clockwise fluid rotation corresponds to positive values of pseudovorticity and vorticity (the ring is moving upward).
  Note that the vorticity and pseudovorticity values in the nozzle proximity, for vertical positions smaller than 15~mm, are not taken into account in the other calculations reported here.}
  \label{omaps}
\end{figure}

Having set these conditions, we calculated the Lagrangian pseudovorticity $\theta$ from the instantaneous positions and velocities of both particle distributions.
In the left and middle panels of figure~\ref{omaps} we display the $\theta$ maps obtained 10~s after the beginning of the simulation.
In the right panel the same-time flow vorticity is plotted.

If one compares the left and right panels of figure~\ref{omaps}, it is apparent that the vortex ring vertical position obtained from the \emph{experiment} particle distribution is quite close to that resulting from the vorticity map.
On the other hand, the ring spatial extent for the E distribution is appreciably larger than that apparent from the vorticity map.
Additionally, the \emph{experiment} pseudovorticity magnitudes are at least 100~times smaller than the vorticity ones.

\begin{figure}
  \centerline{\includegraphics[width = 0.7\linewidth]{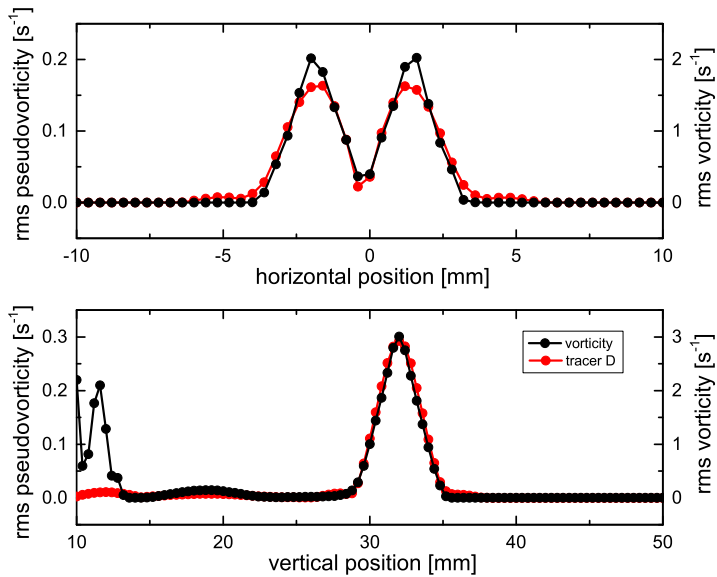}}
  \caption{Root-mean-square pseudovorticity (red) and vorticity (black) profiles obtained 10~s after the beginning of the FEM simulation.
  The pseudovorticity is computed from the \emph{delay} (D) particle distribution.
  Profiles are averaged either in the vertical (top panel) or horizontal (bottom panel) directions.
  The different scales on the vertical axes are for pseudovorticity (left axes) and vorticity (right axes).
  The reference system is as in figure~\ref{omaps}.
  Note that the vorticity and pseudovorticity values in the nozzle proximity, for vertical positions smaller than 15~mm, are not taken into account in the other calculations reported here.}
  \label{oslice}
\end{figure}

Instead, for the \emph{delay} particle distribution the magnitudes of $\theta$ are significantly closer to those of $\omega$ -- see the middle and right panels of figure~\ref{omaps}.
More importantly, the corresponding spatial distributions are quite similar to each other.
This is further confirmed in figure~\ref{oslice}, which shows root-mean-square profiles of $\theta$ and $\omega$, respectively averaged over the vertical and horizontal directions.

Note in passing that particle positions characterised by velocities smaller than 0.01~mm~s$^{-1}$ were not taken into account for the present pseudovorticity estimate because this led to (slightly) smoother $\theta$~maps and (slightly) larger pseudovorticity magnitudes for the \emph{experiment} particle distribution.
The effects were, however, absent for the \emph{delay} particle distribution, due to the larger velocities of the considered particles.
Additionally, for the D distribution, the vorticity in the nozzle proximity is not seen, because the particles are distributed 1~s after the beginning of the simulation.

More generally, the results displayed in figures \ref{omaps} and \ref{oslice} confirm that the Lagrangian pseudovorticity can be used to characterize vortex rings' features in the absence of Eulerian data \citep{pato20}.
Additionally, they demonstrate that the quantitative agreement between $\theta$ and $\omega$ strongly depends not only on the number of flow-probing particles, but also on how they are distributed in the region of interest.



\subsection{Ring position, velocity and radius}
\label{ssec:fem:pos}

In order to further support the previous claims, we calculated other relevant quantities from the obtained pseudovorticity and vorticity maps, which enable, for example, the estimation of the vortex ring location and size over time.

As a first step, following \citet{pato20}, we chose, for each map, a positive threshold value $\theta_0$ that is employed to identify and track the vortex ring.
The positive (clockwise) part of the ring is then made of grid points with $\theta > \theta_0$, the negative (counter-clockwise) one is instead described by $\theta < -\theta_0$.
The chosen values -- listed in table~\ref{Tfem} -- are constant in time and are set to approximately 10~\% of the maximum magnitude at late times.


Once the filtered maps are computed, we can identify the centres of the positive and negative vortical regions, and the distance between them, which is known as the ring diameter $2R$ \citep{gan10}.
Additionally, the ring position $\boldsymbol{P}$ is defined as the centre of the two vortices and its velocity is calculated by convolving $\boldsymbol{P}$ with a suitable Gaussian kernel \citep{pato20}.

Note in passing that the values listed in table~\ref{Tfem} are obtained for times ranging between 3.00 and 14.79~s (the entire simulation lasts 15~s).
The upper limit is due to the Gaussian algorithm chosen for the ring propagation velocity calculation, while the lower one is imposed to remove the influence on the results of the fluid vorticity in the nozzle proximity, see figure~\ref{omaps} -- in the present work we decided not to investigate the ring generation process mainly because this was not accessed in relevant experiments \citep{pato20}.

\begin{table}
  \begin{center}
  \begin{tabular}{lcccccc}
  data set  & $\theta_0$ & $v$             & $v / U_p$ & $2R$            & $C$              & $A_C$            \\[3pt]
  vorticity & 0.800      & 2.08 $\pm$ 0.13 & 0.35      & 3.09 $\pm$ 0.14 & 46.21 $\pm$ 2.32 & 13.78 $\pm$ 0.61 \\
  tracer D  & 0.070      & 2.09 $\pm$ 0.24 & 0.35      & 3.37 $\pm$ 0.09 & 5.37 $\pm$ 0.61  & 16.95 $\pm$ 2.58 \\
  tracer E  & 0.002      & 2.06 $\pm$ 0.42 & 0.34      & 4.51 $\pm$ 0.23 & 0.28 $\pm$ 0.04  & 35.06 $\pm$ 1.91 \\
  \end{tabular}
  \caption{FEM ring properties.
  (i)~Chosen threshold $\theta_0$ for ring identification in s$^{-1}$, set to approximately 10~\% of the maximum magnitude at late times.
  (ii)~Ring vertical velocity $v$ in mm~s$^{-1}$; see also the bottom panel of figure~\ref{oposvel}.
  (iii)~Ratio between the mean value of the ring vertical velocity $v$ and the piston velocity $U_p$, which is found to be consistent with values reported in the literature \citep[see, for example,][]{sullivan08}.
  (iv)~Ring diameter $2R$ in mm; see also figure~\ref{orad}.
  (v)~Ring circulation $C$ in mm$^2$~s$^{-1}$ from \eqref{EC}; see also figure~\ref{ocirc}.
  (vi)~Ring area $A_C$ in mm$^2$; see also figure~\ref{oarea}.
  Results obtained by applying the chosen threshold values.
  Symbols as in \citet{pato20}.
  \label{Tfem}}
  \end{center}
\end{table}

\begin{figure}
  \centerline{\includegraphics[width = 0.7\linewidth]{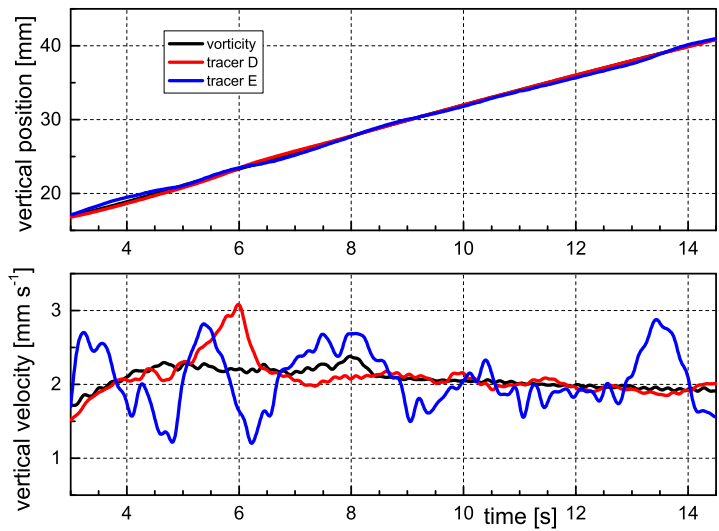}}
  \caption{Ring position (top) and velocity (bottom) as a function of the time $t$ elapsed from the beginning of the FEM simulation.
  The reference system is that defined in figure~\ref{domain}.}
  \label{oposvel}
\end{figure}

In the top (bottom) panel of figure~\ref{oposvel} we plot the position (velocity) of the vortex ring as a function of the time elapsed from the beginning of the simulation.
It is evident that the ring position is well captured by both pseudovorticity maps, while the ring velocity obtained from the \emph{experiment} particle distribution is characterised by a significantly larger standard deviation compared to that derived from the \emph{delay} distribution, see also table~\ref{Tfem}.
On the other hand, the corresponding mean values are quite close to each other, see again table~\ref{Tfem}.
Note also that, at sufficiently late times, the ring velocity shows the tendency to decrease in magnitude, which is a behaviour occurring to vortex rings propagating in a viscous fluid, see, for example, \citet{max74} and \citet{sullivan08}.

\begin{figure}
  \centerline{\includegraphics[width = 0.7\linewidth]{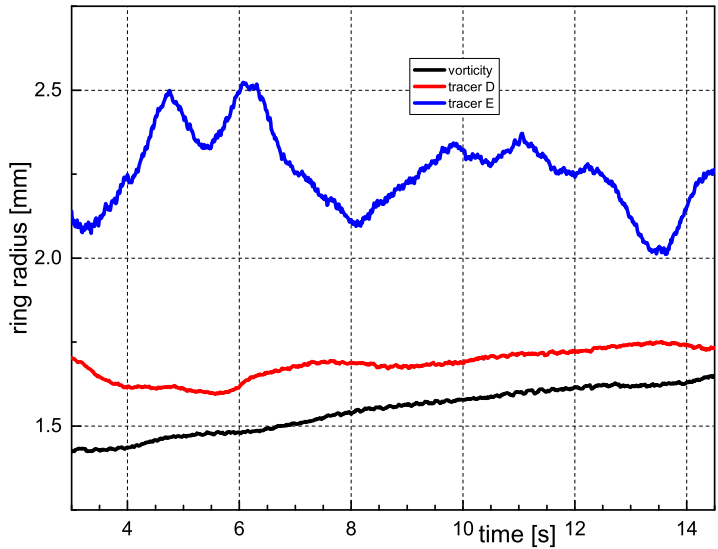}}
  \caption{Ring radius $R$ as a function of the time $t$ elapsed from the beginning of the FEM simulation.
  The reference system is as in figure~\ref{oposvel}.}
  \label{orad}
\end{figure}

One also expects that the decrease of the ring velocity magnitude is accompanied by the increase of its radius.
As shown in figure~\ref{orad}, which displays the vortex radius as a function of the time elapsed from the beginning of the simulation, this behaviour is indeed captured by the \emph{delay} particle distribution.
However, this is not the case for the \emph{experiment} distribution.
The outcome can be related to the above mentioned fact that the ring spatial extent is not well resolved by this particle arrangement, see again figure~\ref{omaps}.


We can therefore say that, by using the Lagrangian pseudovorticity, one can track rather well the position of a vortex ring and estimate with some confidence its propagation velocity.
On the other hand, in order to get quantitative information on the ring spatial extent, one should also employ for the Lagrangian analysis a number of fluid particles significantly larger than those used to date in experiments \citep{pato20}.
Additionally, these particles should be distributed in an relatively small area, with dimensions comparable to those of the ring itself.


\subsection{Ring circulation and area}
\label{ssec:fem:circ}

Following \citet{pato20} we also estimate the ring circulation $C$ from the pseudovorticity (and vorticity) maps.
For this, we use the relation
\begin{equation}
  C = \frac{C_+ - C_-}{2} = \frac{\sum(\theta_+ a) - \sum(\theta_- a)}{2},
  \label{EC}
\end{equation}
where the summations only include the grid points characterised by values of $\theta$ (or $\omega$) passing the threshold $\theta_0$ listed in table \ref{Tfem}.
The subscript $+$ ($-$) denotes that the pseudovorticity and vorticity values are positive (negative), and $a$ is the fixed area associated to each grid point.
Since $C_+ \approx |C_-|$, we discuss here only their mean absolute value $C$.

\begin{figure}
  \centerline{\includegraphics[width = 0.7\linewidth]{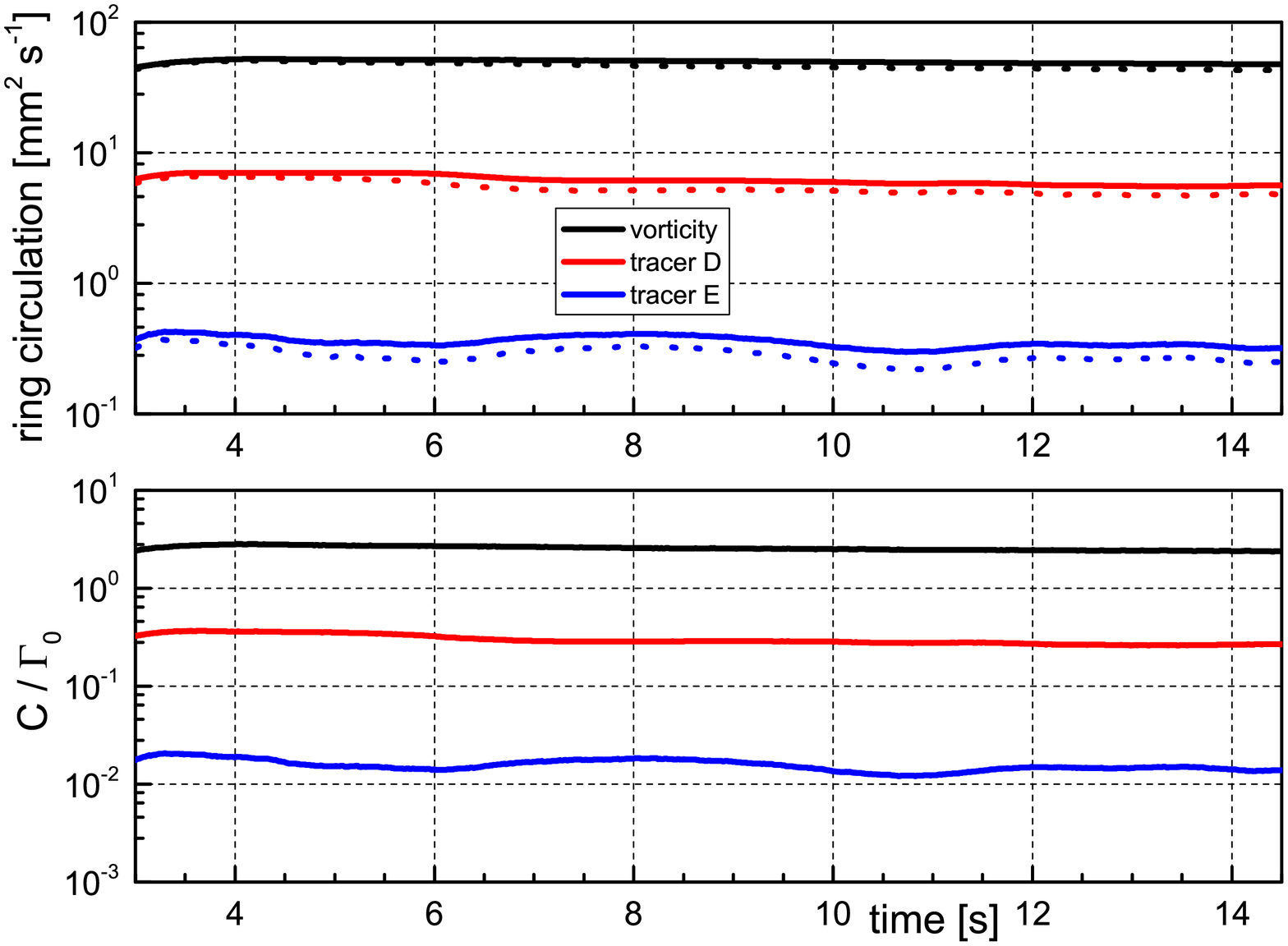}}
  \caption{Ring circulation as a function of the time $t$ elapsed from the beginning of the FEM simulation.
  In the top panel the solid lines indicate the ring circulation $\Gamma$ obtained using the full maps, from (\ref{EC}) with $\theta_0 = 0$, see figure~\ref{omaps}.
  The dashed lines denote the circulation $C$ calculated using the filtered maps, from (\ref{EC}) with the threshold values listed in table~\ref{Tfem}.
  The bottom panel shows $C$ normalised by the ring nominal circulation $\Gamma_0$.
  Note that in the bottom panel and in all the other FEM figures -- excluding figures~\ref{omaps} and \ref{oslice} -- the chosen pseudovorticity and vorticity thresholds are taken into account for the estimate of the plotted quantities.}
  \label{ocirc}
\end{figure}

The temporal evolution of $C$ is represented in the top panel of figure~\ref{ocirc} (dashed lines) and, in the same panel, the solid lines correspond to the actual ring circulation $\Gamma$, obtained using the full maps, that is, with the threshold $\theta_0$ set to zero.
We can see that $C$ is consistently smaller than $\Gamma$, likely because the filtered maps do not capture the wake shed by the moving ring.
The panel also displays the tendency of the ring circulation to decrease at late times.
This is an expected behaviour for both turbulent and laminar rings, as discussed, for example, by \citet{max74} and \citet{didden79}, respectively.

Additionally, the circulation magnitude derived from the flow vorticity (black line) is approximately 100~times larger than that obtained from the \emph{experiment} particle distribution (blue line).
This is also evident from the bottom panel of figure~\ref{ocirc}, which displays $C$ normalised by the ring nominal circulation $\Gamma_0$.
A similar outcome was reported by \citet{pato20}, who showed that experimentally obtained values of $C$ can be much smaller than relevant values of $\Gamma_0$, up to approximately two orders of magnitude.
Note that for these experiments it was not possible to access the actual ring circulation $\Gamma$ and that the nominal circulation was estimated from suitable values of fluid velocity.

From the bottom panel of figure~\ref{ocirc} it is also apparent that $C / \Gamma_0 \approx 2$, when $C$ is computed from the filtered vorticity maps (black line).
Indeed, measured values of ring circulation were found to be of the same order of $\Gamma_0$ by a few investigators, in various experimental conditions \citep[see, for example,][]{max74,didden79,borner85}.
Note, however, that the ring nominal circulation $\Gamma_0 = U_p^2 t_p / 2$; that is, if we set, for our FEM ring, $U_p = 10$~mm~s$^{-1}$ and $t_p = 0.8$~s, we obtain a nominal circulation more than two times larger than that employed in the bottom panel of figure~\ref{ocirc} -- see also the above discussion on \eqref{Ere}.
It then follows that the numerical values shown in the panel should solely be regarded as first-order estimates, considering also that, as discussed, for example, by \citet{glezer88}, the nozzle geometry and the piston velocity time history may have a significant influence on the resulting ring features.

\begin{figure}
  \centerline{\includegraphics[width = 0.7\linewidth]{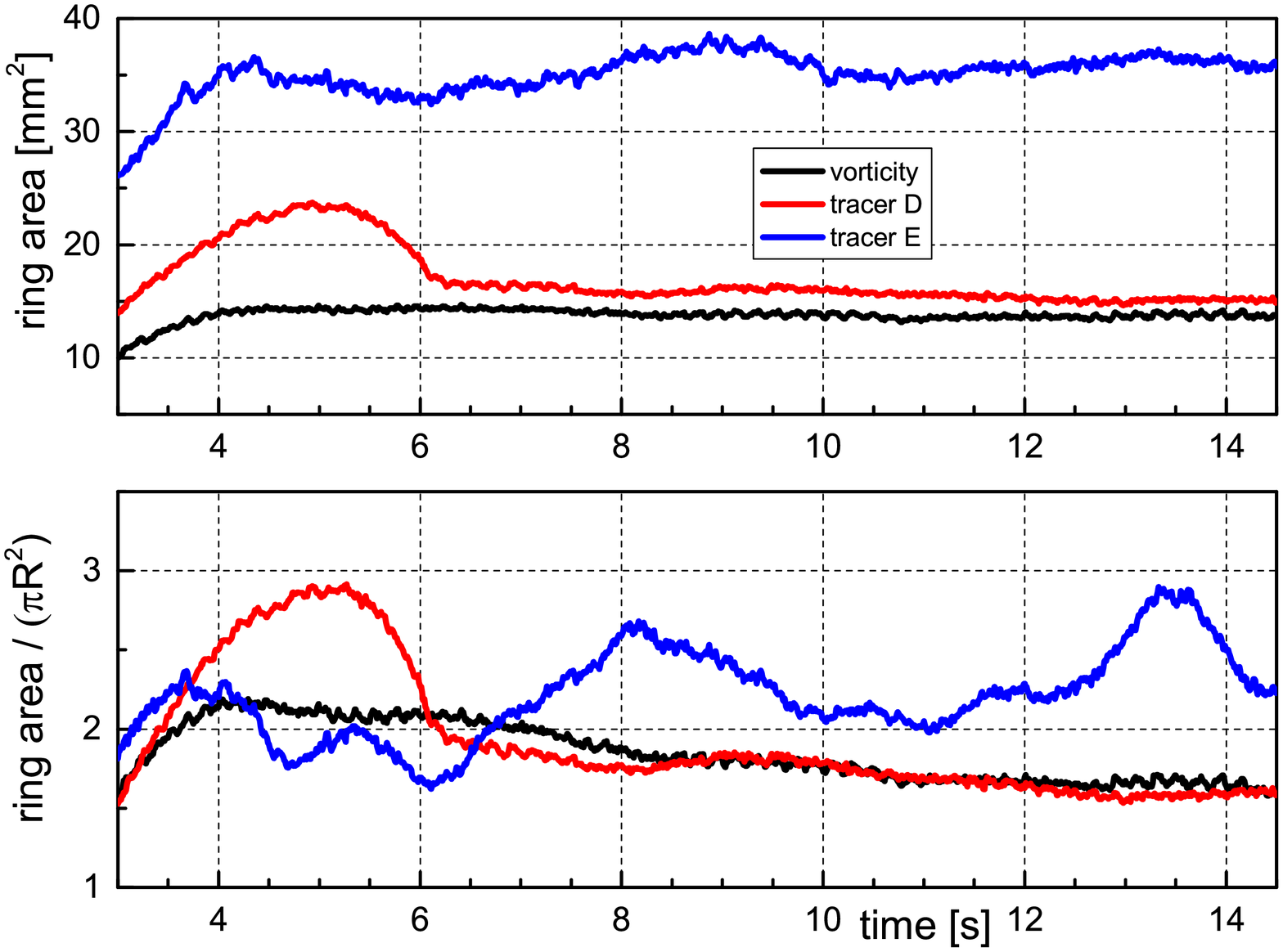}}
  \caption{Ring area as a function of the time $t$ elapsed from the beginning of the FEM simulation.}
  \label{oarea}
\end{figure}

Following \citet{pato20} we can now estimate the ring area $A_C$ from \eqref{EC}, by setting $\theta_+ = 1$ and $\theta_- = -1$.
We plot the result in the top panel of figure~\ref{oarea}, as a function of the time elapsed from the beginning of the simulation.
Additionally, in the bottom panel, we display the ratio $A_C / A_R$, where $A_R = \pi R^2$ and $R$ is the estimated ring radius shown in figure~\ref{orad}.

The figure most prominent feature is likely the peak observed for the \emph{delay} particle distribution (red lines) at short times.
This can be related to the fact that, at around the same time, the ring wake disappears from the filtered $\theta$ maps.
Note that a similar peak is also observed in the estimated vertical velocity of the ring, at $t \approx 6$~s (bottom panel of figure~\ref{oposvel}).

Additionally, from the bottom panel of figure~\ref{oarea}, it appears that the ring radius $R$ can be used as a valuable first-order estimate of the ring spatial extent.
More importantly, we see from both panels that the ring spatial extent is quite well captured by the \emph{delay} particle distribution, consistently with our previous discussion.

In summary, our results confirm that the Lagrangian pseudovorticity can be used to quantify general features of relatively large vortical structures -- such as their position and propagation velocity -- and that this is especially meaningful when Eulerian data are not available, which is specifically the case of experiments in superfluid $^4$He \citep{pato20}.
Additionally, we have shown here that the spatial extent of these vortices can be quantitatively determined by $\theta$ maps solely in conditions yet to be met in relevant experiments, that is, only for sufficiently dense particle distributions, which, by the way, could allow to directly access Eulerian flow fields.

So far we have considered fluid particles, which by definition exactly follow the flow, but the particles commonly used in experiments can deviate from streamlines due to their inertia.
In the next section, we then investigate the effect of particle inertia on the relation between $\theta$ and $\omega$.


\section{HVBK ring}
\label{sec:hvbk}

The Hall--Vinen--Bekarevich--Khalatnikov (HVBK) model is often used to describe the large-scale hydrodynamics of superfluid $^4$He.
At finite temperatures, between approximately 1 and 2.2~K, this unique liquid can be seen as a mixture of two components: the superfluid component, with no viscosity, and the normal fluid component, with Newtonian viscosity \citep[see, for example,][]{Landau1941,donnelly09}.
The HVBK model is valid at scales much larger than the typical distance between quantised vortices, which are line singularities -- holes -- within the superfluid component.
The model therefore describes the coarse-grained dynamics of the two components.
Within this framework, their respective velocity fields $\vs$ and $\vn$ follow two coupled incompressible Navier--Stokes equations,
\begin{align}
  \label{eq:HVBK_vn}
  \diffp{\vn}{t} + (\vn \vdot \gradient) \vn
  &= -\frac{\gradient \pn}{\rhon} + \nun \laplacian \vn - \frac{\rhos}{\rhon} \Fns,
  \\
  \label{eq:HVBK_vs}
  \diffp{\vs}{t} + (\vs \vdot \gradient) \vs
  &= -\frac{\gradient \ps}{\rhos} + \nus \laplacian \vs + \Fns,
  \\
  \label{eq:HVBK_div}
  \gradient \vdot \vs &= \gradient \vdot \vn = 0.
\end{align}
Here, $\rhos$ and $\rhon$ are the temperature-dependent densities of the two fluids (the total fluid density is $\rho = \rhon + \rhos$), while $\ps$ and $\pn$ indicate their respective pressure fields.

The two fluids are coupled by a mutual friction force $\Fns$.
This force is responsible for the transfer of energy between the components, and it also contributes to the dissipation of energy of the entire system.
As a first approximation, $\Fns$ increases linearly with the relative velocity between the components, that is, $\Fns = \alpha \OmegaMF (\vn - \vs)$, where $\alpha$ is a non-dimensional mutual friction coefficient -- tabulated by \citet{donnelly98} -- and $\OmegaMF$ indicates a mutual friction frequency, related to the density and orientation of the quantised vortices embedded in the superfluid.
For example, in isotropic superfluid turbulence, this frequency may be taken as proportional to a characteristic superfluid vorticity \citep{Lvov2006}.
Here, as we consider the case of an isolated vortex ring, this frequency is taken as a constant control parameter.
Note also that we have included in \eqref{eq:HVBK_vs} a superfluid dissipation term, parameterised by an effective temperature-dependent kinematic viscosity $\nus$ \citep{Vinen2002}.
This term accounts for the dissipation of superfluid energy at scales smaller than those resolved by the HVBK model.
Such dissipation mechanisms include quantised vortex reconnections and excitation of Kelvin waves, which transport energy towards the smallest scales of the system \citep[see, for example,][]{svancara2019,villois}.
Additionally, in \eqref{eq:HVBK_vn} the kinematic viscosity of the normal component $\nun$ is set equal  to $\mu_\text{n} / \rhon$, where $\mu_\text{n}$ is the fluid dynamic viscosity, also tabulated by \citet{donnelly98}.

In the reference study by \citet{pato20}, solid deuterium particles were used to evaluate the Lagrangian pseudovorticity associated to vortex rings generated in superfluid \HeFour{}.
Such particles are not perfect tracers of the flow, as their density is about $1.4$ times larger than that of the fluid, and their typical diameter is a few micrometres \citep{grid}.
Hence, particle inertia is expected to play some role on pseudovorticity estimates.

In superfluid \HeFour{}, inertial particles are not only subject to the Stokes drag associated to the normal fluid viscosity, but also respond to pressure gradient forces from the two fluids \citep[see, for example,][]{Sergeev2009}.
Finite size effects can be neglected, as particles are much smaller than the core size of the studied vortex ring, as reported below, in \S\ref{ssec:hvbk:simulations}.
Consequently, if one also neglects Basset history terms and assumes that particles are spherical, the particle equation of motion reads
\begin{equation}
  \label{eq:HVBK_particles}
  \diff{\vp}{t} =
  \frac{\vn(\xp) - \vp}{\StokesTime} +
  \beta \left(
    \frac{\rhon}{\rho} \Diff{\vn}{t} +
    \frac{\rhos}{\rho} \Diff{\vs}{t}
  \right),
\end{equation}
where $\xp$ and $\vp$ are the particle position and velocity, respectively, and $\mathrm{D} / \mathrm{D}t$ denotes the material derivative associated to each fluid component.
The density parameter $\beta$ accounts for added mass effects and is given by $\beta = 3\rho / (2\rhop + \rho)$, where $\rhop$ is the particle density.
The Stokes time, representing the typical particle response time to normal fluid fluctuations, is $\StokesTime = \ap^2 / (3 \beta \nu)$, where $\ap$ denotes the particle radius, and $\nu = (\rhon / \rho) \nun = \mu_\text{n} / \rho$ indicates the temperature-dependent kinematic viscosity of superfluid \HeFour{}, also tabulated by \citet{donnelly98}.
Formally, tracer dynamics is obtained in the limit $\StokesTime\to0$.

The HVBK equations~\eqref{eq:HVBK_vn}--\eqref{eq:HVBK_div} are solved by direct numerical simulation in a three-dimensional periodic domain using a Fourier pseudo-spectral solver \citep{Homann2009}.
Particles are initialised at random locations in the domain and are advanced in time according to \eqref{eq:HVBK_particles}.
Fluid velocities and accelerations are interpolated at particle positions using fourth-order B-splines \citep{vanHinsberg2012}.
Time advancement of fields and particles is performed using a third-order Runge--Kutta scheme.

\begin{figure}
  \centerline{\includegraphics[width = 0.7\linewidth]{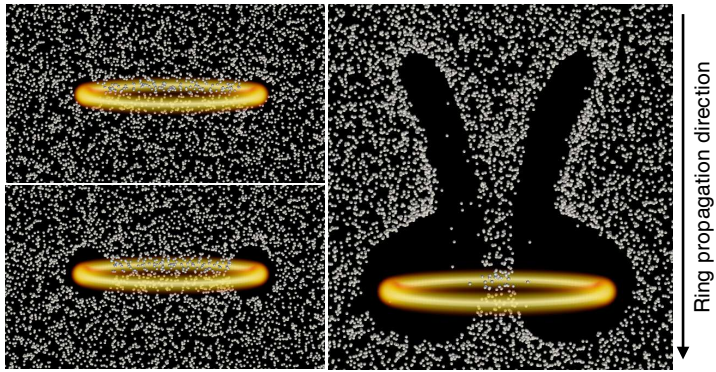}}
  \caption{HVBK simulation of a vortex ring moving downward with three families of deuterium particles at a relatively early time, approximately 1~s after the beginning of the simulation.
  Top left: particles with radius of $5\,\mu$m.
  Bottom left: particles with radius of $10\,\mu$m.
  Right: particles with radius of $30\,\mu $m.
  The respective Stokes numbers are $St \approx 0.1$, $0.3$ and $2.5$ -- see table~\ref{Thvbk} in \S\ref{ssec:hvbk:pos}.
  The peculiar shape observed in the right panel is due to particles being expelled from highly rotating regions as a result of their high inertia.
  Additionally, particles initially located in front of the vortex ring are propelled upwards via the ring centre in a ballistic motion, resulting in the observed wake pattern.}
  \label{gholes}
\end{figure}

Figure~\ref{gholes} displays results from a typical simulation, in which we observe the vortex ring together with spherical particles having the same density of deuterium particles, at the chosen temperature $T = 1.9\,$K (see numerical details later).
All panels correspond to the same time of ring evolution, for different particle families.
From left to right the particle size is increased, leading to a larger Stokes response time.
As a consequence, the effect of inertia becomes apparent and holes (regions without particles) are clearly observed.
Such structures are similar to those seen in the experimental pictures displayed in figure~\ref{holes}, but one should keep in mind that the latter are not Eulerian images, that is, they were not obtained at a fixed time as the pictures in figure~\ref{gholes}.

\subsection{Simulation setup}
\label{ssec:hvbk:simulations}

We consider superfluid $^4$He at 1.9\,K.
At this temperature, the densities of the normal fluid and superfluid components satisfy the ratio $\rhon/\rhos=0.72$ and the liquid kinematic viscosity is $\nu \approx 0.01$~mm$^2$~s$^{-1}$.
This value is very close to that used for the FEM ring -- note that, as stated above, the density and viscosity of He~II depend on temperature \citep[see, for example,][]{donnelly98}.
Additionally, at this temperature, the effective superfluid viscosity is close to the normal fluid one \citep{Vinen2002} and the mutual friction coefficient is $\alpha=0.206$. For the numerical simulations we set the mutual friction parameter to $\Omega_0\approx110$~s$^{-1}$.
We can thus consider that the fluid components are well locked to each other for the time scales relevant to this work.


An isolated vortex ring of normal fluid is introduced in the middle of the periodic box, which has 28\,mm sides.
The origin of the chosen reference system is located at the box centre.
The ring radius is $R = 3.5\,\text{mm}$ and its initial circulation is $\Gamma_0 \approx 95$\,mm$^2$~s$^{-1}$.
It follows from \eqref{Ere} that the ring Reynolds number is slightly larger than $10^4$.
Note that the superfluid component is initially at rest.
However, due to mutual friction, the normal fluid vortex ring quickly induces the formation of an equivalent superfluid ring.
Such a macroscopic superfluid vortex is typically composed of a bundle of quantised vortices \citep[see, for example,][]{wacks} that are modelled in the HVBK framework as a continuous vorticity field.
In this work we are not interested in this fast transient dynamics.

Note also that, due to the domain periodicity, a vortex ring eventually returns to its initial position.
Although the ring leaves a wake behind it, its dynamics is almost unaffected by it, since the wake is rapidly dissipated by viscosity (within one period, lasting about 5~seconds).
However, as particles are not continuously injected in the domain, their spatial distribution becomes inhomogeneous after the passing of a ring.
Hence, the studied configuration mimics an experiment in which rings are injected periodically, but one should also account for the fact that the ring size and velocity change with time.
We will not further exploit this analogy in the present work.

The vortex ring core is initialised with a Gaussian vorticity profile, $\omega(r) = \omega_0 \exp \left[ -r^2 / (2 r_0^2) \right]$, where $r$ is the distance from the core centre.
The vortex core radius $r_0 = 0.2\,\text{mm}$ and the vorticity at the core centre is set to fix the value of the ring circulation.
Note in passing that the generation of such a thin vortex ring does not appear straightforward with the apparatus employed by \citet{pato20}, mainly because the experimentally generated rings seem to have a rather large core size, but, as shown below, the HVBK ring provides a good setting to study the effect of inertia.

As detailed in \S\ref{ssec:hvbk:pseudo}, we track solid hydrogen and deuterium particles -- by setting their density -- with sizes similar to those found in \HeFour{} experiments.
For comparison, we also follow perfect tracers of the normal fluid.
We specifically consider three families of spherical particles, having radii of 5, 10 and 30\,$\mu$m, and we seed $5 \times 10^5$ particles, homogeneously distributed over the whole (three-dimensional) computational domain.
The corresponding Stokes times range approximately from 1 to 41~ms.
The characteristic ring time $\tau_\text{r}$ is estimated using the initial circulation $\Gamma_0$ and the core radius $r_0$.
We obtain that $\tau_\text{r} \approx 17$~ms, i.e. relevant values of Stokes number $St =  \StokesTime / \tau_\text{r}$ are between $0.04$ and $2.47$ -- see also table~\ref{Thvbk} in \S\ref{ssec:hvbk:pos}.

\subsection{Pseudovorticity maps}
\label{ssec:hvbk:pseudo}

At the beginning of the simulation, particles are suspended in the cubic volume.
A layer of the latter, parallel to the ring propagation direction and situated in the centre of the periodic box, is considered for the pseudovorticity estimate.
This volume layer is 0.2~mm thick and contains between 3100 and 3600 particles at each time step, which lasts approximately 0.09~s (the entire simulation lasts approximately 27~s).
The positions and velocities of these particles were then employed for the calculation of the pseudovorticity according to \eqref{ET}.
This estimation was performed on a square grid of $101 \times 101$ inspection points, covering a $27.6 \times 27.6\,\text{mm}^2$ area.
The chosen annular region has 0.1\,mm inner and 3\,mm outer radii, which are the same values used for our FEM analysis in a viscous fluid, discussed above, in \S\ref{sec:fem}.
The corresponding time window is equal to the time step, that is, for the HVBK ring, no temporal averaging was carried out.

Note also that, as in the FEM study, the numerically obtained flow vorticity values were interpolated on the chosen pseudovorticity grid, for the sake of comparison.
Additionally, in order to follow the ring motion for distances larger than the box size, the ring is at each time step centred in the box middle by using the position of the normal fluid vorticity maximum at that time step (the introduced ring is symmetric).
From the latter maximum position it is then possible to get the ring position as a function of time, in a chosen reference system not moving with the ring, as for the FEM ring.

Pseudovorticity maps were obtained for fluid tracers and inertial particles.
The latter are characterised by a density different from that of the fluid and by a finite Stokes response time (a finite radius).
Additionally, we chose particles having features compatible with recent visualisation experiments in superfluid $^4$He \citep[see, for example,][]{prf}.
Specifically, we chose spherical particles made of solid deuterium $D_2$ -- heavier than liquid He~II, with a density ratio of $1.37$ at $1.9$~K -- and solid hydrogen $H_2$ -- lighter than superfluid $^4$He, with a density ratio of $0.60$ at the same temperature.
Similarly, we selected particle radii of the same order of and larger than the dimensions of particles in experiments -- see \citet{epl} and \citet{grid} for typical distributions of particle sizes.

\begin{figure}
  \centerline{\includegraphics[width = 1\linewidth]{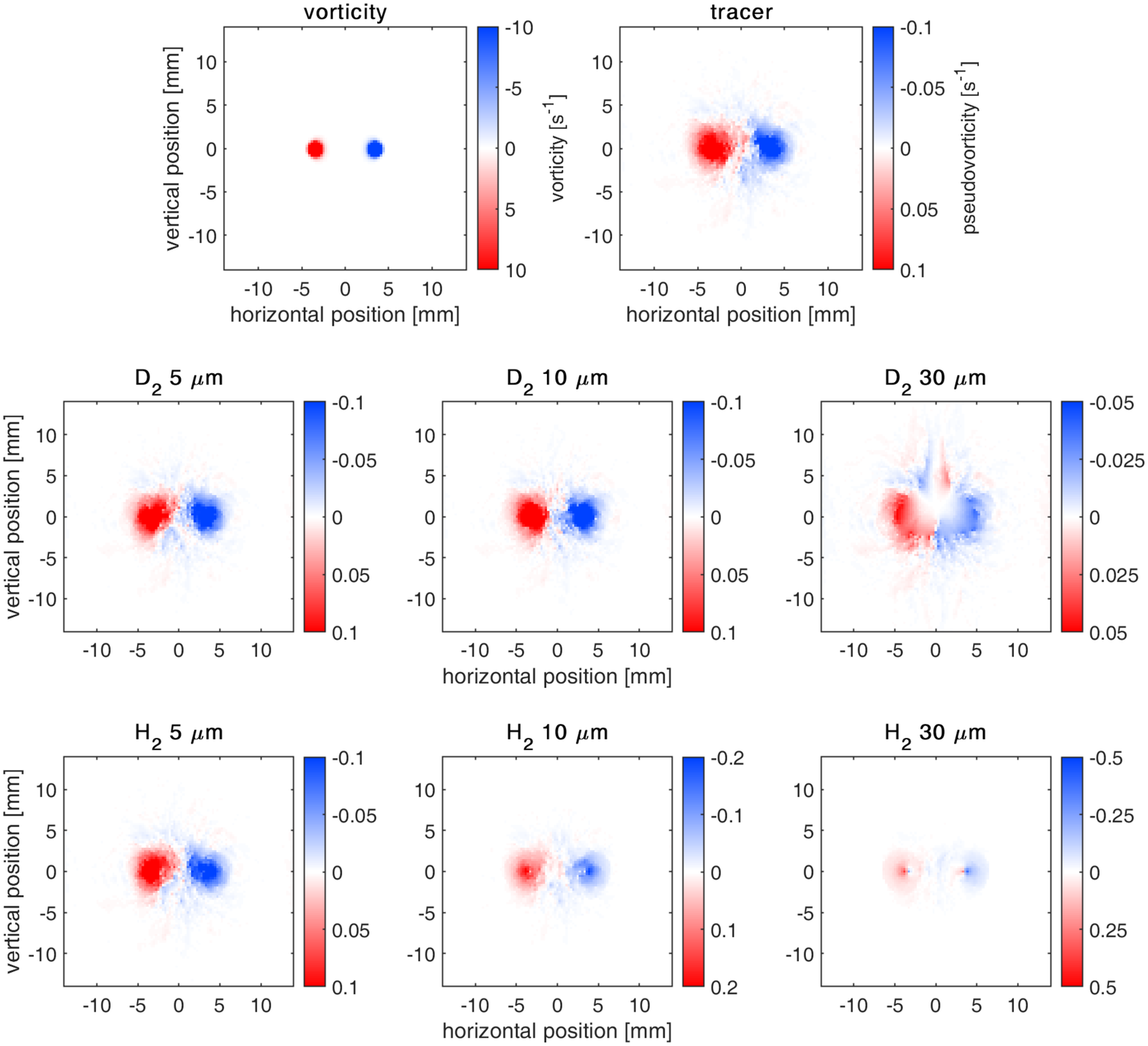}}
  \caption{Vorticity and pseudovorticity maps obtained 2.69~s after the beginning of the HVBK simulation.
  The top left panel displays the Eulerian vorticity map.
  The other panels show pseudovorticity maps obtained using different types of particles, as indicated in the panel titles (the shown dimension indicates the particle radius).
  The  origin of the reference system, moving with the ring, is located at the centre of the periodic box.
  Clockwise fluid rotation corresponds to positive values of pseudovorticity and vorticity (the ring is moving downward).}
  \label{gmaps}
\end{figure}

Pseudovorticity maps calculated from the different particle classes are shown in figure~\ref{gmaps}.
As in the FEM analysis, it is possible to capture the instantaneous ring location by using tracer particles, but the magnitude of the computed pseudovorticity is two orders of magnitude smaller than that of the vorticity; see also the corresponding circulation values in table~\ref{Thvbk}, in \S\ref{ssec:hvbk:pos}.
Moreover, the pseudovorticity maps obtained using small inertial particles are quantitatively similar to those obtained using tracers.
This is expected, as the motion of small particles should not be strongly affected by inertial effects.
More practically, it indicates that hydrogen and deuterium particles having radii up to 10\,$\mu$m -- corresponding to  $St < 0.3$ -- may be used to accurately detect the presence of a vortex ring in the present setting.
In our study, it is only for the largest particles, with radius $\ap = 30$\,$\mu$m, corresponding to $St > 1$, that pseudovorticity estimates importantly deviate from those of tracer particles, even though relevant qualitative information may still be extracted from the resulting maps.

In our simulations, we also observe that, for fluid tracers, the number of particles does not influence appreciably the obtained results.
For example, if the chosen layer of the cubic volume is five times thicker, the number of particles used for the pseudovorticity calculation also increases by approximately five times, but the corresponding magnitude does not change appreciably, although its behaviour in space and time is possibly smoother than that obtained for the thinner layer.
Note that similar results were obtained for the FEM ring, that is, increasing the number of particles in a given area did not significantly improve the agreement between vorticity and pseudovorticity magnitudes above a certain number of particles, see also the inset of figure~\ref{oseen}.
We expect that, for given number of particles, a better agreement could be obtained for smaller annuli, as discussed in appendix \S\ref{sec:anal}.
On the other hand, if one decreases the size of the region used for the pseudovorticity estimate in actual experiments, the number of particles in that region also decreases \citep{pato20}.
It then follows that, as mentioned above, the choice we made here for the annulus size is mainly motivated by the requirement of having enough particles in the region of interest for the pseudovorticity calculation.


\begin{figure}
  \centerline{\includegraphics[width = 1\linewidth]{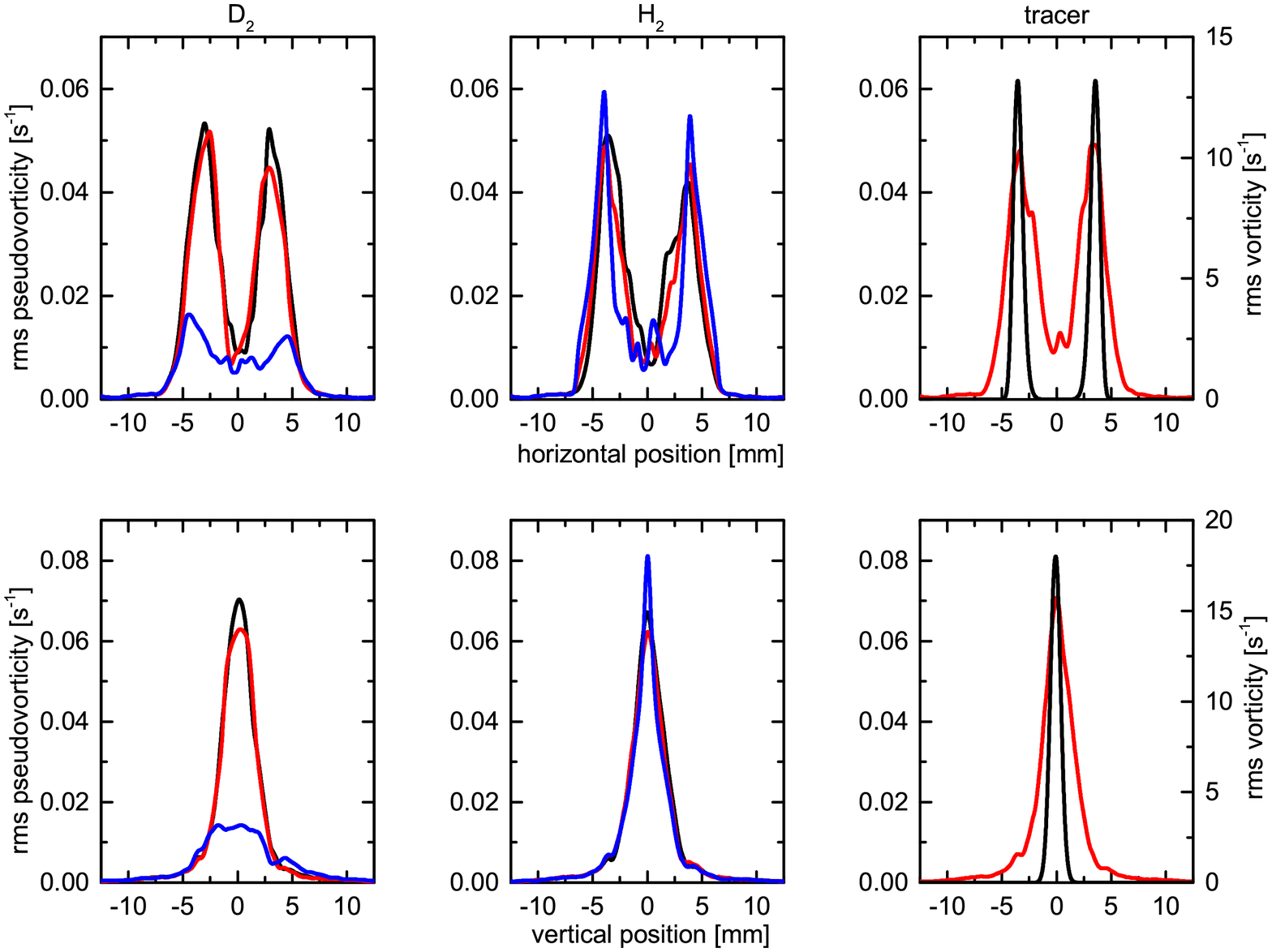}}
  \caption{Root-mean-square pseudovorticity and vorticity profiles obtained 2.69\,s after the beginning of the HVBK simulation.
  Profiles are averaged either in the vertical (top panels) or horizontal (bottom panels) directions.
  Left and middle panels display pseudovorticity profiles obtained from deuterium, $D_2$, and hydrogen, $H_2$, particles, respectively.
  Black, red and blue lines in these panels indicate that corresponding particle radii are 5, 10 and 30 $\mu$m, respectively.
  On the right panels pseudovorticity profiles for tracers are plotted as red lines.
  Black lines on the same panels indicate corresponding vorticity profiles.
  Note the different scales on the vertical axes of the right panels for pseudovorticity (left axes) and vorticity (right axes).
  The reference system is as in figure~\ref{gmaps}.}
  \label{gslice}
\end{figure}

In order to make a quantitative comparison, we present in figure~\ref{gslice} the root-mean-square profiles of pseudovorticity for the different particle families.
These are compared with the corresponding vorticity profiles.
We clearly observe that particle inertia becomes more important for the heavier and larger particles, i.e. for the deuterium particles with 30\,$\mu$m radii, corresponding to the largest Stokes number.
Additionally, vorticity and pseudovorticity profiles qualitatively differ, in agreement with the above discussion.


\subsection{Ring position, velocity and radius}
\label{ssec:hvbk:pos}

Table~\ref{Thvbk} lists relevant vortex ring properties averaged over the entire HVBK simulation, for times ranging from 0.09 to 26.93\,s.
The exception is the ring propagation velocity, which is averaged between 1.89 and 25.14\,s due to the Gaussian algorithm chosen for its calculation \citep{pato20}.
Note that the little dependence of the ring velocity on the data set is most likely related to the above mentioned procedure of ring positioning in the box centre at each time step.
Indeed, this centering procedure was performed by using the normal fluid vorticity maximum, whose magnitude and position do not depend on the particle type.

\begin{table}
  \begin{center}
  \begin{tabular}{lcccccc}
  data set        & $\theta_0$ & $v$             & $2R$            & $C$              & $A_C$            & $St$ \\[3pt]
  vorticity       & 1.000      & $5.24~\pm$ 0.77 & 7.10 $\pm$ 0.11 & 82.86 $\pm$ 5.30 & 11.63 $\pm$ 4.22 & --   \\
  tracer          & 0.010      & $5.24~\pm$ 0.78 & 6.53 $\pm$ 0.25 & 1.51 $\pm$ 0.16  & 35.10 $\pm$ 2.96 & --   \\
  D$_2$~5~$\mu$m  & 0.010      & $5.24~\pm$ 0.78 & 6.36 $\pm$ 0.21 & 1.58 $\pm$ 0.14  & 34.70 $\pm$ 2.51 & 0.07 \\
  D$_2$~10~$\mu$m & 0.010      & $5.24~\pm$ 0.77 & 5.79 $\pm$ 0.17 & 1.74 $\pm$ 0.10  & 37.37 $\pm$ 3.04 & 0.27 \\
  D$_2$~30~$\mu$m & 0.005      & $5.28~\pm$ 0.84 & 4.99 $\pm$ 0.47 & 0.74 $\pm$ 0.30  & 51.52 $\pm$ 3.34 & 2.47 \\
  H$_2$~5~$\mu$m  & 0.010      & $5.24~\pm$ 0.78 & 6.63 $\pm$ 0.28 & 1.49 $\pm$ 0.16  & 36.58 $\pm$ 3.13 & 0.04 \\
  H$_2$~10~$\mu$m & 0.020      & $5.24~\pm$ 0.77 & 7.32 $\pm$ 0.44 & 1.24 $\pm$ 0.18  & 25.18 $\pm$ 1.79 & 0.16 \\
  H$_2$~30~$\mu$m & 0.050      & $5.24~\pm$ 0.77 & 6.34 $\pm$ 1.14 & 1.60 $\pm$ 0.37  & 14.24 $\pm$ 2.63 & 1.45 \\
  \end{tabular}
  \caption{HVBK ring properties.
  (i)~Chosen threshold $\theta_0$ for ring identification in s$^{-1}$, set to approximately 10~\% of the maximum magnitude at late times.
  (ii)~Ring vertical velocity $v$ in mm~s$^{-1}$; see also the bottom panels of figure~\ref{gposvel}.
  (iii)~Ring diameter $2R$ in mm; see also figure~\ref{grad}.
  (iv)~Ring circulation $C$ in mm$^2$~s$^{-1}$ from \eqref{EC}.
  (v)~Ring area $A_C$ in mm$^2$.
  (vi)~Stokes number $St =  \StokesTime / \tau_\text{r}$, where $\tau_\text{r} \approx 17$~ms.
  Results obtained by applying the chosen threshold values.
  Symbols as in \citet{pato20}.
  \label{Thvbk}}
  \end{center}
\end{table}

\begin{figure}
  \centerline{\includegraphics[width = 1\linewidth]{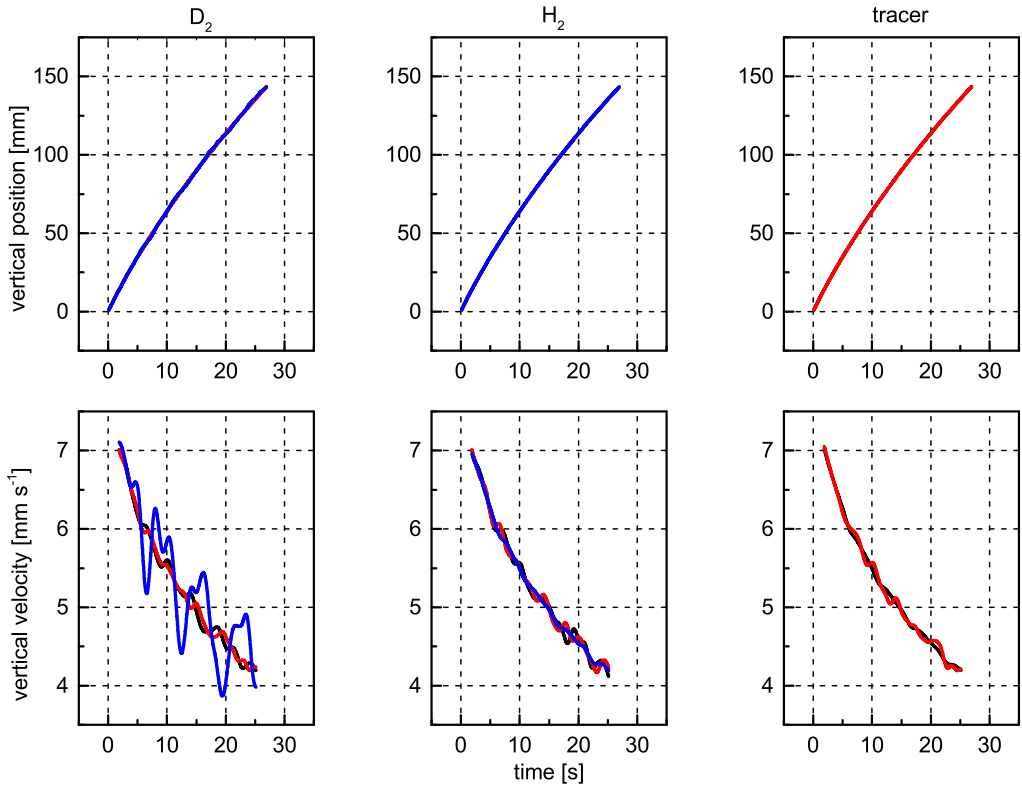}}
  \caption{Temporal evolution of the ring vertical position (top) and velocity (bottom) obtained from the HVBK simulation.
  Panel disposition and line colour code as in figure~\ref{gslice}.
  Note that here, for the sake of clarity and comparison, we plot the absolute values of ring position and velocity in a fixed coordinate system, not moving with the ring and coinciding with the moving frame of reference at the beginning of the simulation (in the coordinate system of figure~\ref{gmaps} the ring is moving downward, in the negative direction).
  The velocity time limits are due to the Gaussian algorithm chosen for the velocity calculation \citep{pato20}.}
  \label{gposvel}
\end{figure}

In figure~\ref{gposvel} we present the vortex ring vertical position and velocity, computed from the pseudovorticity maps.
For comparison, we also show the results directly obtained form the vorticity field, plotted as black lines in the right panels.
Remarkably, the determination of the ring location is robust and independent of the particle family.
Concerning its velocity, the general trend is well reproduced, but some periodic oscillations are visible for the heaviest and larger particles -- namely the $D_2$ $30\,\mu$m family, corresponding to the largest Stokes number.
These oscillations are likely related to the periodicity of the domain because their period -- about 5\,s -- is approximately the time it takes for the ring to travel through the periodic box.
Additionally, as mentioned above, when the ring passes, it leaves in its wake an inhomogeneous distribution of particles that is later encountered in front of the ring.
This effect is more marked for particles with high inertia, as it is also apparent from figure~\ref{gholes}.

\begin{figure}
  \centerline{\includegraphics[width = 0.7\linewidth]{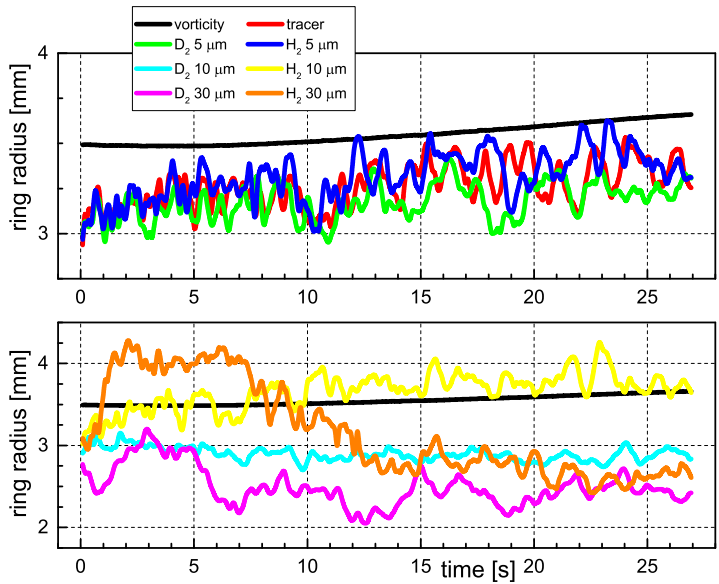}}
  \caption{Ring radius as a function of the time $t$ elapsed from the beginning of the HVBK simulation.
  Note the different scales on the vertical axes.
  Reference system as in figure~\ref{gposvel}.}
  \label{grad}
\end{figure}

The determination of the ring radius, plotted in figure~\ref{grad}, is more sensitive to particle inertia.
It is evident from the figure -- see especially the top panel -- that the ring radius estimate does not depend appreciably on particle size and density for small enough particles, with $St \lesssim 0.1$.
Additionally, for these particles, the expected time increase of the radius, evident from the vorticity trend, is recovered; see also the related discussion in \S\ref{ssec:fem:pos}.

Another remarkable feature that can be seen in figure~\ref{grad} is that, for particle radii of 5 and 10~$\mu$m, the ring radius $R$ is -- on average -- larger for the hydrogen particles than for the heavier ones (the ring diameter $2R$ was defined above, in \S\ref{ssec:fem:pos}, as the distance between the observed counter-rotating vortices).
This may indicate that deuterium particles -- heavier than the fluid -- tend not to probe the ring core region, as visualised in experiments (figure~\ref{holes}) and in numerical simulations (figure~\ref{gholes}).
Moreover, for deuterium particles with radii equal to 10 and 30~$\mu$m, the ring radius does not change appreciably with time and is consistently smaller than the tracer one.


The peculiar behaviour observed in the bottom panel of figure~\ref{grad} for the largest hydrogen particles, those with $St = 1.45$,  is at present not entirely understood.
However, from the corresponding pseudovorticity maps -- not shown here -- one can notice that, for sufficiently large times, larger than approximately 10~s, the underlying dipolar ring structure is replaced by two adjacent dipoles.
On the other hand, one may more generally say that such large particles are not suitable for tracking the ring.

Note finally that, for the sake of completeness, we also list in table~\ref{Thvbk} the values of ring circulation and area for the different particle families, in the same fashion as in table~\ref{Tfem}, but we do not show here the corresponding time trends because they do not add any information on our discussion, that is, they merely confirm the already reported findings.


\section{Conclusions}
\label{sec:conc}

In a recent experimental work \citep{pato20} particle tracking velocimetry was used to construct Lagrangian pseudovorticity maps that were found to provide valuable information on the propagation of macroscopic vortex rings in superfluid $^4$He.
In that work, it was also shown analytically that such maps are closely related to the Eulerian vorticity.
For such a mathematical proof, it was assumed that particles act as ideal fluid tracers and that they are homogeneously distributed in space.
In practice, such assumptions are rarely met in experiments.

In this work, we have numerically studied how particles can be used to study the propagation of vortex rings, in the spirit of \citet{pato20}.
We have performed two sets of numerical simulations.
In the first set, we have employed the FEM method to study the behaviour of an isolated vortex ring in a viscous fluid, in a geometry similar to the experimental setting used by \citet{pato20}.
Fluid tracers were specifically added to the flow to investigate the reliability of pseudovorticity maps.
In the second set, we have performed numerical simulations of an isolated vortex ring in superfluid $^4$He, in a periodic domain, using the HVBK model.
In this case, we have considered the dynamics of several particle families of different sizes and densities.
We have thus addressed the effect of particle inertia on the propagation properties of an isolated vortex ring.

Overall, our results confirm that the Lagrangian pseudovorticity is a relevant estimator for quantifying general features of isolated vortical structures, such as their position and propagation velocity.
This observation is robust and independent of particle inertia.
However, we have seen that the intensity of pseudovorticity fields can be much weaker than the actual Eulerian vorticity, especially for sparse particle distributions, and that the ring size is not correctly captured when highly inertial particles are used.

The outcome is especially important in the absence of Eulerian data, which is the case of state-of-the-art visualization experiments in superfluid $^4$He \citep[see, for example,][]{pato21}.
More generally, our work neatly demonstrates that quantitative comparisons between particle trajectories obtained in different experimental conditions can be performed by using pseudovorticity maps, taking also into account particle concentration and inertia.
Additionally, future work could focus on establishing the relevance of such a measure for the identification of Lagrangian coherent structures, which is especially relevant for geophysical flows \citep[see, for example,][]{hadjighasem}.\\

\textbf{Acknowledgements.} We thank P.~\v{S}van\v{c}ara for valuable help.
HVBK computations were carried out on the M\'esocentre SIGAMM hosted at the Observatoire de la C\^ote d'Azur and on the French HPC cluster OCCIGEN through the GENCI allocation A0072A11003.
J.~I.~Polanco and G.~Krstulovic were supported by the Agence Nationale de la Recherche through the GIANTE ANR-18-CE30-0020-01 project.
O.~Outrata, M.~Pavelka, J.~Hron and M.~La~Mantia acknowledge the support of the Czech Science Foundation (GA\v{C}R) under grant no.~19-00939S.\\

\textbf{Declaration of Interests.} The authors report no conflict of interest.


\appendix
\section{Analytical results}
\label{sec:anal}

\citet{pato20} reported that the Lagrangian pseudovorticity $\theta$ is half of the Eulerian vorticity $\omega$. This specifically occurs in the close vicinity of the inspection point $\rr$, defined above, \eqref{ET}, and for homogeneous and isotropic distributions of fluid particles, having smooth velocity fields.
Although these conditions are at present not met in the case of experiments in He~II, the result clearly demonstrates that $\theta$ and $\omega$ are related, and therefore supports the use of the Lagrangian pseudovorticity, especially in the absence of Eulerian data.

In this section we further investigate analytically the relation between pseudovorticity and vorticity in order to explain some results reported in the main text.

Following \cite{pato20} we first rewrite (\ref{ET}) as
\begin{equation}
  \theta(\rr,R) = \frac{1}{N} \int_{D(\rr,R)} f(\rr') \frac{\left[\left(\rr' - \rr\right) \times \uu(\rr')\right]_z} {|\rr' - \rr|^2} \dd^2 \rr',
  \label{ETf1}
\end{equation}
where $D(\rr,R)$ is a circle, centred at $\rr$, with radius $R$,
\begin{equation}
  N = \int_{D(\rr,R)} f(\rr') \dd^2 \rr',
  \label{ETf2}
\end{equation}
and $f$ denotes the particle distribution function (note that here we disregard the time dependence of the problem).

We will now estimate the pseudovorticity for a two-dimensional vortex characterised by a purely azimuthal fluid velocity, having magnitude
\begin{equation}
  v_a = \frac{\Gamma}{2\pi r} g(r^2),
  \label{EAvel}
\end{equation}
where $\Gamma$ is the constant fluid circulation associated to the vortex and $r$ indicates the distance from the vortex centre.
We assume here that $g(r^2)$ is a smooth, non-negative function, vanishing at the origin, $g(r^2) = \OBig(r^2)$ for $r \rightarrow 0$, and at most constant at infinity, $g(r^2) = \OBig(1)$ for $r \rightarrow \infty$.
The corresponding flow vorticity magnitude can then be obtained as
\begin{equation}
  \omega = \frac{1}{r} \frac{\partial r v_a}{\partial r} =\frac{\Gamma}{\pi} g'(r^2),
  \label{EAvort}
\end{equation}
where $g'(r^2)$ indicates the relevant derivative of $g(r^2)$.

The pseudovorticity for a circle of radius $R$, centred at the inspection point $(x_0,0)$, in the Cartesian reference system having the origin at the vortex centre, can then be analytically determined, for a given function $g(r^2)$, if also the spatial particle distribution $f$ is known.
We will first work with an unknown function $g$ and an uniform particle distribution $f = N/(\pi R^2)$, $N$ being the number of fluid particles.
By using (\ref{EAvel}) we can then rewrite (\ref{ETf1}) as
\begin{equation}
  \theta(x_0;R) = \frac{1}{\pi R^2} \int_{D(x_0,R)} \frac{\rr - (x_0,0)}{|\rr - (x_0,0)|^2} \times \left[\frac{\Gamma}{2\pi r} g(r^2) \ee_a \right]\dd^2 \rr,
  \label{ETx0R}
\end{equation}
where $\ee_a$ indicates the azimuthal velocity direction in the chosen Cartesian coordinate system, and $r$ is the magnitude of $\rr$.

In the polar reference system having the origin at the inspection point $(x_0,0)$ (\ref{ETx0R}) becomes
\begin{equation}
  \theta(x_0;R) = \frac{\Gamma}{2\pi^2 R^2} \int_0^R \int_0^{2\pi} \frac{g(r^2)}{r^2} \left(r'+ x_0 \cos\varphi'\right) \dd\varphi'\dd r'
  \label{ETx0Rp}
\end{equation}
where $r'$ and $\varphi'$ are the chosen polar coordinates, and $r^2 = (r')^2 + 2 r' x_0  \cos\varphi' + x^2_0$.
Since near the origin $g(r^2) = \OBig(r^2) = g'(0) r^2 + \OBig(r^4)$, the integrand in (\ref{ETx0Rp}) has no singularity; $g'(0)$ denotes here the derivative of $g(r^2)$ at the vortex centre.
Additionally, for $x_0 \rightarrow \infty$ the integral in (\ref{ETx0Rp}) goes to zero, that is, the pseudovorticity vanishes far from the vortex centre, because the velocity vanishes too.

For $x_0 = 0$, when the inspection point $(x_0,0)$ coincides with the vortex centre, (\ref{ETx0Rp}) becomes
\begin{equation}
  \theta(0;R) = \frac{\Gamma}{\pi R^2} \int_0^R \frac{g[(r')^2]}{r'}\dd r'.
  \label{ET0Rp}
\end{equation}
Since $g$ behaves at infinity at most as a constant, the integral in (\ref{ET0Rp}) behaves at most as $\ln R$.
Therefore, for large $R$ values, $\theta(0;R)$ goes to zero.
For small $R$ values, we obtain
\begin{equation}
  \lim_{R \rightarrow 0} \theta(0;R) =
  \lim_{R \rightarrow 0} \frac{\Gamma}{\pi R^2} \int_0^R \frac{g'(0) (r')^2 + \OBig[(r')^4]}{r'}\dd r'
  = \frac{\Gamma}{2\pi} g'(0) = \frac{\omega|_{r=0}}{2},
  \label{ET0R0}
\end{equation}
which means that, at the origin, the pseudovorticity converges to half of the vorticity.
This is compatible with the general result reported by \citet{pato20}, where it was shown that, for particle distributions radially symmetric around the inspection point, the pseudovorticity converges to half of the vorticity, for sufficiently small $R$ values.

Additionally, by using the Hermite-Hadamard inequality \citep{Niculescu}, it can be shown that $\theta(0;R)$ is a decreasing function of $R$ near the origin, because the function $g(r^2)/r$ is there concave.
However, for large $R$ values, the latter must become convex, since it is positive, but, at the same time, it is expected that, for large $R$ values, the pseudovorticity goes to zero and therefore its variations are negligible.

\begin{figure}
  \centerline{\includegraphics[width = 0.7\linewidth]{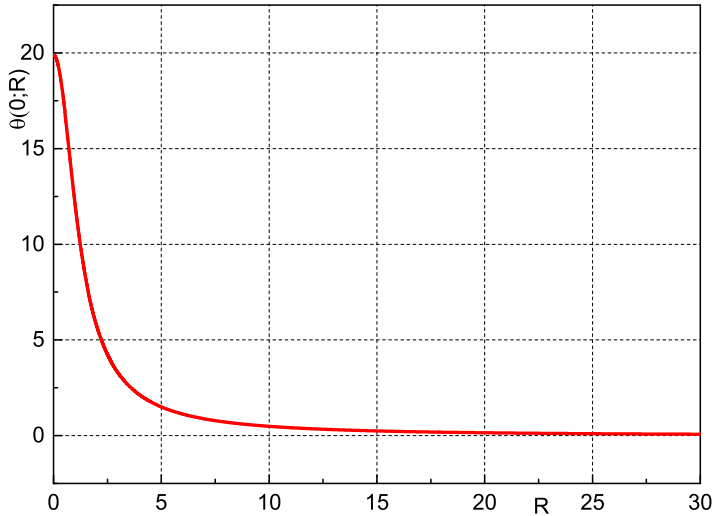}}
  \caption{Pseudovorticity $\theta(0;R)$ as a function of the circle radius $R$, (\ref{ET0Rp}), for a Lamb-Oseen vortex, (\ref{ELOvel}), with $\Gamma = 50$, $t = 10$ and $\nu = 0.01$.
  Note that the pseudovorticity $\theta(0;R)$ approaches half of the vorticity, $\omega/2 \approx 20$, (\ref{ELOvort}), for $R \rightarrow 0$, vanishes for $R\rightarrow \infty$ and is a decreasing function of $R$.
  For the sake of simplicity and generality, we do not explicitly mention in this section relevant physical units.}
  \label{mplor}
\end{figure}

In figure~\ref{mplor} we plot $\theta(0;R)$ as a function of $R$, for a Lamb-Oseen vortex, having the azimuthal velocity magnitude
\begin{equation}
  v_{LO} = \frac{\Gamma}{2\pi r} g(r^2) = \frac{\Gamma}{2\pi r} \left[1 - \exp \left(\frac{-r^2}{4t\nu}\right)\right],
  \label{ELOvel}
\end{equation}
where $r$ denote the distance from the vortex centre and, as in figure~\ref{oseen}, we set $\Gamma = 50$, $t = 10$ and $\nu = 0.01$ (for the sake of simplicity and generality, we do not explicitly mention in this section relevant physical units).
Note in passing that the corresponding flow vorticity magnitude can be written as
\begin{equation}
  \omega_{LO} = \frac{\Gamma}{4\pi t\nu} \exp \left(\frac{-r^2}{4t\nu}\right).
  \label{ELOvort}
\end{equation}

In summary, the analytical results reported to date -- see especially figure~\ref{mplor} -- supports the view that the pseudovorticity magnitude becomes smaller if the circular area chosen for its estimate increases, as it is also mentioned in the main text.

\begin{figure}
  \centerline{\includegraphics[width = 0.7\linewidth]{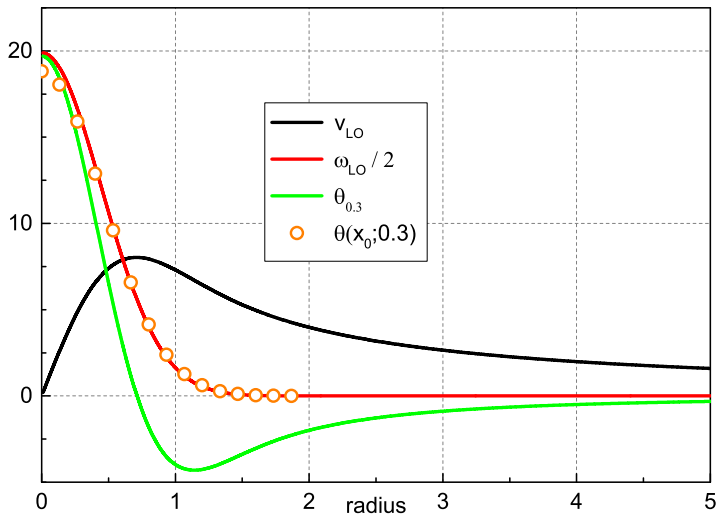}}
  \caption{Pseudovorticity for a Lamb-Oseen vortex as a function of the distance from the vortex axis (radius).
  Vortex parameters as in figures \ref{oseen} and \ref{mplor}, that is, $\Gamma = 50$, $t = 10$ and $\nu = 0.01$.
  Black line: vortex azimuthal velocity $v_{LO}$, \eqref{ELOvel}; red line: half of the vorticity $\omega_{LO} / 2$, \eqref{ELOvort}; green line: pseudovorticity corresponding to the open green circles in the main panel of figure~\ref{oseen}, estimated as described in that figure caption, that is, by using directly \eqref{ET}; open orange circles: pseudovorticity $\theta(x_0;R)$, \eqref{ETx0Rp}, with $R = 0.3$.}
  \label{mplor1}
\end{figure}

For $x_0 \neq 0$ the pseudovorticity is given by \eqref{ETx0Rp} and we numerically calculated the integral for the just introduced Lamb-Oseen vortex, the one with $\Gamma = 50$, $t = 10$ and $\nu = 0.01$.
The integration was carried out by using the software Wolfram Mathematica and in figure~\ref{mplor1} we plot the obtained result for $R = 0.3$ (open orange circles).
A remarkable agreement with half of the actual flow vorticity (red line) is obtained.

In the same figure the green line, showing a pseudovorticity sign change, corresponds to the green open circles in the main panel of figure~\ref{oseen}, that is, it displays the pseudovorticity for the chosen Lamb-Oseen vortex, estimated directly from \eqref{ET}, by using 1000 equally spaced grid points, up to a radius of 10.
Additionally, 100 equally spaced fluid particles were distributed in the chosen annular region, having outer radius equal to $0.3$ (the pseudovorticity was not computed in the region centre and was set to zero at the vortex centre).
Note also that, as mentioned in \S\ref{sec:pseudo}, the particles are in this case equally spaced along the radial direction, which means that the chosen particle distribution is not radially symmetric and that the particles are clustered in the vortex centre vicinity.

We can see from \eqref{ETx0Rp} that the integrand is non-negative if
\begin{equation}
  r' + x_0 \cos\varphi' \geq 0,
  \label{ETn1}
\end{equation}
which can be rewritten in the Cartesian coordinate system $(x,y)$ having the origin at the vortex centre as
\begin{equation}
  \left(x - \frac{x_0}{2} \right)^2 + y^2 \geq \left(\frac{x_0}{2}\right)^2.
  \label{ETn2}
\end{equation}
The above equality is satisfied for the circle with radius $x_0/2$, centred at $(x_0/2,0)$.
Points outside this circle contribute positively to the pseudovorticity, while points inside it contribute negatively, and the intersection of $D(x_0,R)$ with the positively contributing region is larger than that with the negatively contributing region.
Therefore, if the function $g(r^2)/r^2$ behaves nearly as a constant within $D(x_0,R)$, the contribution of the positive region (outside that circle) will be larger than that of the negative region (within that circle), and the pseudovorticity will be positive.
However, if the function $g(r^2)/r^2$ is considerably larger in the negative region, the pseudovorticity can become negative.
Since the function $g(r^2)/r^2$ is here assumed to be smooth, the pseudovorticity should typically be positive, as indicated in figure~\ref{mplor1} (open orange circles).

It then follows that the sign changes of the pseudovorticity displayed in figures \ref{oseen} and \ref{mplor} -- see also the middle panel of figure~\ref{omaps} -- are most likely an effect of the discrete particle distributions employed for the $\theta$ estimates based on \eqref{ET}, especially if one also considers that the analytical results reported here are only valid for uniform particle distributions, that is, we set $f = N/(\pi R^2)$ in \eqref{ETf1}.
Indeed, we would like to address analytically the effect of particle distribution non-uniformities in future studies.

Another possible line of future research could be based on the following result, which can be considered a generalisation of the analytical relation between $\theta$ and $\omega$ reported by \citet{pato20}.
The latter was obtained by expanding the fluid velocity to the second order with respect to the difference $(\rr' - \rr)$.
If, in a similar fashion, we expand it to the fourth order, we get
\begin{equation}
  \theta(\rr;R) = \frac{\omega(\rr)}{2} + \frac{R^2}{32} \Delta \omega(\rr)
  \label{ET4}
\end{equation}
for a smooth velocity field, in the close vicinity of the inspection point $\rr$, for radially symmetric particle distributions.
The pseudovorticity is thus approximately a quadratic function of $R$ and its behaviour depends on the Laplacian of the vorticity at the inspection point.
It then follows that, if $\theta(\rr;R)$ is known, one could estimate the flow vorticity $\omega(\rr)$ by using \eqref{ET4}.
The latter estimate could be performed by numerical means, or even analytically, and this could be the focus of future studies, willing to assess in more detail the effect of $R$ on the relation between $\theta$ and $\omega$.



\end{document}